\documentclass[12pt]{article}
\usepackage{epsfig,amsfonts,amssymb}
\usepackage{cite}
\input epsf.sty
\def\ZZZ{{\hbox{ Z\kern-1.6mm Z}}}
\def\zzz{{\hbox{z\kern-1mm z}}}

\newcommand{\ws}{{\wt\sigma}}
\newcommand{\wrh}{{\wt\rho}}
\newcommand{\wv}{{\wt v}}

\newcommand{\BB}{{\cal B}}

\newcommand{\AAA}{{\cal A}}

\newcommand{\MM}{{\cal M}}
\newcommand{\CC}{{\cal C}}

\newcommand{\OO}{{\cal O}}

\newcommand{\LL}{{\cal L}}

\newcommand{\wt}{\widetilde}
\newcommand{\wh}{\widehat}

\newcommand{\RR}{{\cal R}}
\newcommand{\NN}{{\cal N}}

\newcommand{\be}{\begin{equation}}
\newcommand{\ee}{\end{equation}}
\newcommand{\ben}{\begin{eqnarray}\displaystyle}
\newcommand{\een}{\end{eqnarray}}

\newcommand{\bea}[1]{\begin{eqnarray}\label{#1} }
\newcommand{\eea}{\end{eqnarray}}

\newcommand{\refb}[1]{(\ref{#1})}

\newcommand{\sectiono}[1]{\section{#1}\setcounter{equation}{0}}

\def\one{{\hbox{ 1\kern-.8mm l}}}
\def\zero{{\hbox{ 0\kern-1.5mm 0}}}

\begin{document}

{}~
{}~
\hfill\vbox{\hbox{hep-th/yymmnnn}
}\break

\vskip .6cm

{\baselineskip20pt
\begin{center}
{\Large \bf Walls of Marginal Stability and 
Dyon Spectrum in $\NN=4$ Supersymmetric String Theories
} 

\end{center} }

\vskip .6cm
\medskip

\vspace*{4.0ex}

\centerline{\large \rm
Ashoke Sen}

\vspace*{4.0ex}

\centerline{\large \it Harish-Chandra Research Institute}

\centerline{\large \it  Chhatnag Road, Jhusi,
Allahabad 211019, INDIA}

\vspace*{1.0ex}

\centerline{\it and}

\vspace*{1.0ex}

\centerline{\large \it Department of Physics, California Institute
of Technology}

\centerline{\large \it Pasadena, CA91125, USA}

\vspace*{1.0ex}

\centerline{E-mail:
sen@mri.ernet.in, ashokesen1999@gmail.com}

\vspace*{5.0ex}

\centerline{\bf Abstract} \bigskip

The spectrum of quarter BPS dyons in 
$\NN=4$ supersymmetric string theories can change as the
asymptotic moduli cross
walls of marginal stability on which
the dyon can break apart into a pair of half BPS states. 
In this paper we
classify these marginal stability walls and
examine this phenomenon in the context of exact dyon spectrum
found in a class of $\NN=4$ supersymmetric string theories.
We argue that the dyon partition functions
in different domains separated by marginal stability walls are the
same, but the choice of integration contour needed
for extracting the degeneracies from the partition function differ
in these different regions. We also find that in the limit of large
charges the change in the degeneracy 
is exponentially suppressed compared to the
leading contribution. This is consistent with the fact that in the
computation of black hole entropy we do not encounter any change
as the asymptotic moduli fields move across the walls of marginal
stability. Finally we carry out some tests of S-duality invariance
in the theory.

\vfill \eject

\baselineskip=18pt

\tableofcontents

\sectiono{Introduction and Summary} \label{sintro}

In a series of papers 
we computed the exact degeneracy of quarter
BPS dyons in a
class of $\NN=4$ supersymmetric string compactifications in
an appropriate corner of the moduli space of these 
theories\cite{0605210,0607155,0609109}
verifying and generalizing
earlier conjectures\cite{9607026,0510147,0602254}.
Alternative approaches to this problem leading to similar results
have also been 
developed\cite{9607026,0505094,0506249,0612011,0603066}. 
The result for the degeneracy takes the form of integration over a
three real
dimensional subspace (a contour)
of the Siegel upper half plane parametrizing
genus two Riemann surfaces, 
and the integrand involves inverse of a certain meromorphic
modular form of a subgroup of the Siegel modular group. 

It is well known however that for a dyon with a given set of
charges the moduli space of $\NN=4$ supersymmetric string theory
contains subspaces on which the original dyon becomes 
marginally unstable
against decay into a pair of other dyons\cite{9712211,9804160}.
In particular the moduli space contains
walls of marginal stability  -- 
codimension one subspaces --
on which the  mass of the quarter BPS dyon becomes equal
to the sum of masses of a pair of half BPS dyons whose charges
add up to that of the original quarter BPS dyon. On this subspace
the original dyon becomes marginally unstable against decay into
this pair of half BPS dyons, and typically the spectrum of the original
dyon changes discontinuously as we move through these marginal
stability walls in the moduli space\cite{9712211,9804160}.
Thus an important question is: how does the dyon spectrum
computed in \cite{0605210,0607155,0609109} change as we
move away from the particular corner of the moduli space
in which the degeneracy was computed?

A glimpse of this issue was already seen in the analysis of
\cite{0605210,0607155,0609109} where it was found that even
in the corner of the moduli space where the result was computed,
-- in a weakly coupled type IIB string theory compactified on
a certain orbifold, -- the result for the degeneracy changes
discontinuously as the angle between certain pair of
circles of the compact manifold
passes through zero. 
This change could be attributed to the fact that precisely at this
point the system under consideration became marginally stable
against decay into a pair of half BPS states.
However the change was such that
the expression for the degeneracy continues to be given by a
similar integral with identical integrand, but the contour over which
the integral is to be performed gets changed. If we try to 
deform the new contour into the original contour we encounter
a pole of the integrand and hence the two contributions differ by
the residue at the pole. One can also interpret the  result by saying
that the dyon partition function formally remains the same as we
move through the marginal stability wall, but the point in the
Siegel moduli space around which we should series expand the 
partition function to extract the degeneracies changes as we move
through these walls.

\begin{figure}
\leavevmode
\begin{center}
\hbox{
\epsfysize=3cm \epsfbox{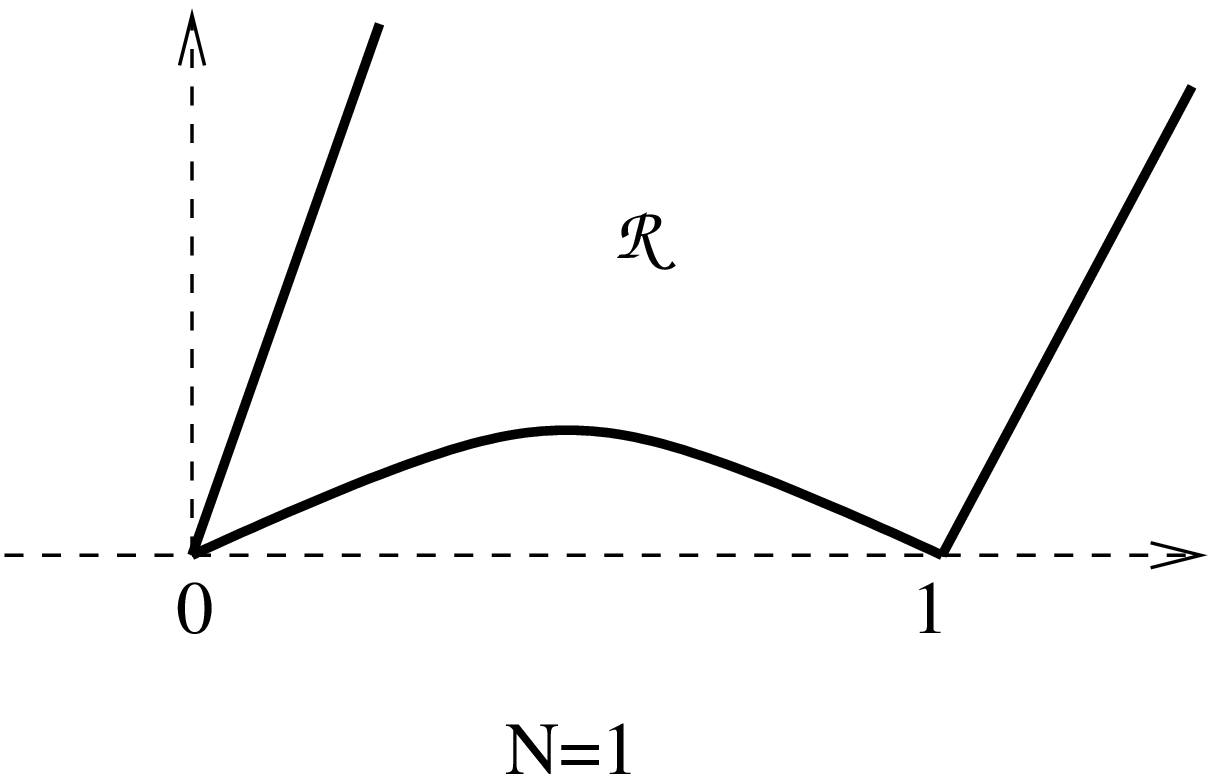}
\epsfysize=3cm \epsfbox{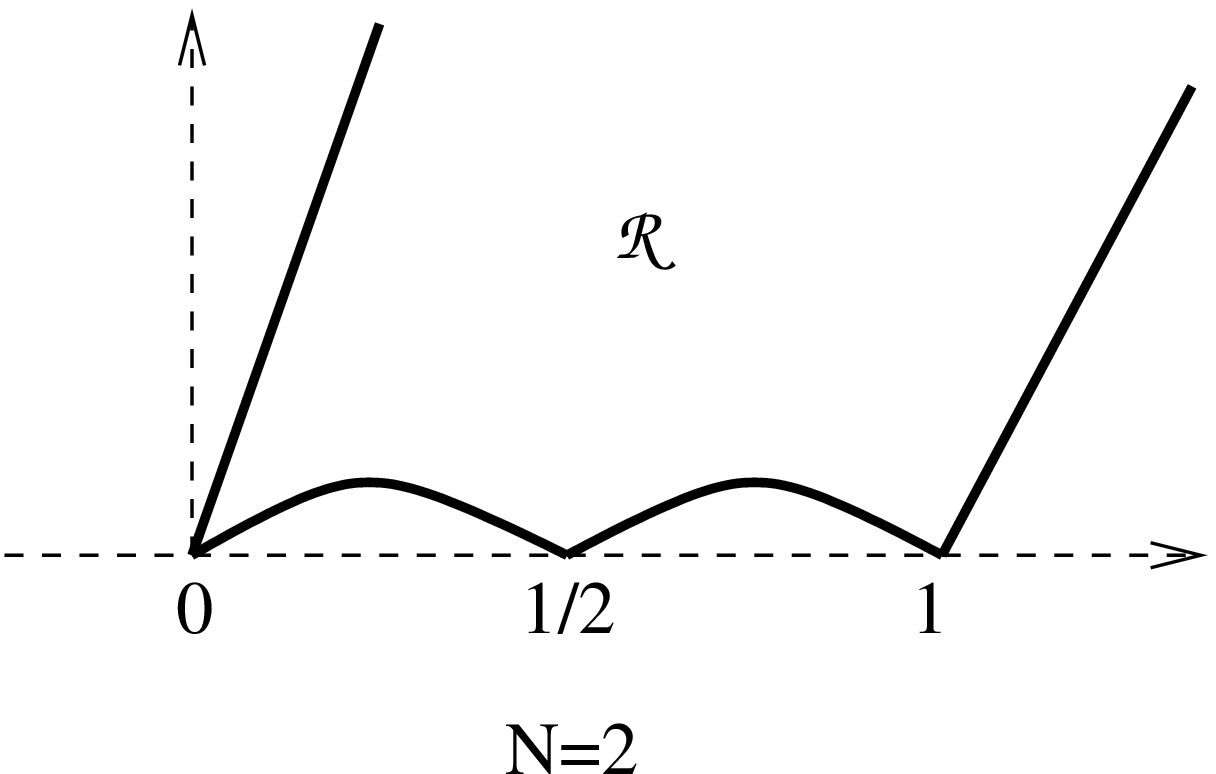}
\epsfysize=3cm \epsfbox{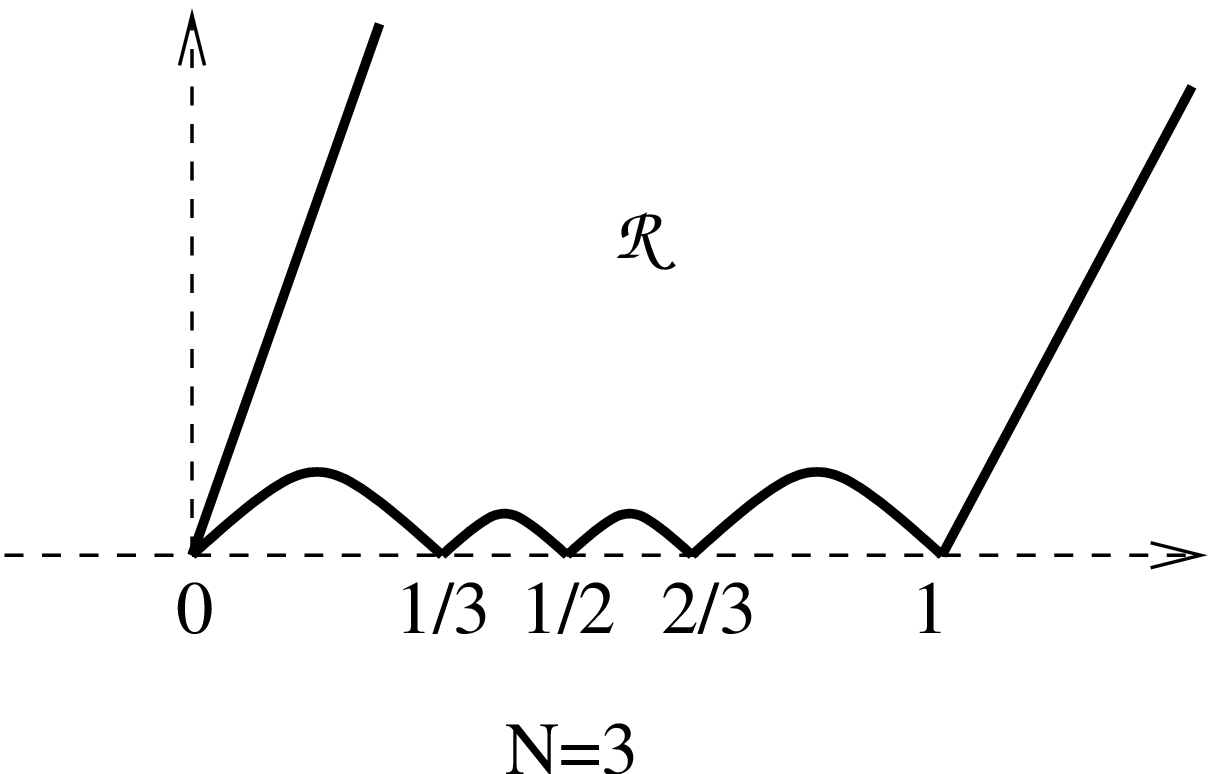}
}
\end{center}
\caption{A schematic diagram representing the domain $\RR$
in the upper half $\tau_\infty$ plane, bounded by the walls of
marginal stability,  for $\ZZZ_1$, $\ZZZ_2$ and $\ZZZ_3$ orbifolds. 
The shapes of the circles and
the slopes of the straight lines bordering the domain depend
on the charges and other asymptotic moduli, but the vertices
are universal.} \label{f1}
\end{figure}

In this paper we classify these marginal stability walls in the
moduli space  
of the theory and explore what happens when we move across these
walls.
Although the walls are complicated codimension one surfaces
in the moduli space, one gets a simpler picture by
regarding them as curves in the
axion-dilaton moduli space for fixed values of the other moduli.
We find that these curves are circles and straight lines, and 
could intersect on the real axis or at $i\infty$, 
but have no intersection in the
interior of the upper half plane. 
Furthermore although the slopes of the lines and the radii and the
centres of the circles depend on the charges and other moduli,
the points where they intersect the real axis and each other are
universal.
As a result a region bounded
by these curves has universal vertices but boundaries which depend
on the other moduli and charges.  One such region has been
displayed in Fig.\ref{f1} in \S\ref{smarginal}.
As we move from one of these regions to another, we cross marginal
stability walls and as a result the dyon spectrum could change.

Typically most of these walls lie outside the domain in which the
approximation made in our 
computation of the degeneracy can be trusted. However
using T- and S-duality invariance of the theory we can extract
useful information about these walls. A useful input in reaching this conclusion
is the observation made in \cite{appear} that under an S-duality
transformation the degeneracy formula does not remain invariant, but
there is a change in the integration contour. This can be attributed to
the fact that a duality transformation acts both on the charges as well
as the asymptotic values of the moduli fields, whereas the degeneracy
formula is computed for different charges but in the same region
of the moduli space, in a sense that will be made precise in
\S\ref{sdualtwo}. The apparent lack of duality symmetry of the 
degeneracy formula is due to the fact that the region in which the
degeneracy formula is calculated and the region obtained after a
duality transformation are typically separated by walls of marginal
stability. Thus the knowledge of how the contour changes under a
duality transformation can be used to extract information about
how it changes from one region of the moduli space to another as
we move through these marginal stability walls.
This way the change in the
spectrum across any
wall of marginal stability can be encoded as the result of changing the
integration contour leaving the integrand unchanged, or equivalently
as shifting the point around which we expand the partition function
to extract the degeneracy. 

One can try to find a physical interpretation of these changes in the
degeneracy by explicitly evaluating the residues at the poles picked
up by the contour as we move it from the position associated
with one side of a wall of marginal stability to that on the other side
of the wall. We can do this easily in the context of the marginal
stability wall found in the original analysis of \cite{0605210}.
We find that the change in the degeneracy is proportional to the
total degeneracy of the half-BPS
electric and magnetic states into which the
original dyon can decay on this particular marginal stability wall.

Given these results, it is natural to ask what 
happens in the large charge
limit where the statistical entropy given by the logarithm of
the degeneracy can be compared with the black hole entropy. We find
that in this limit the change in the degeneracy as we move across a
wall of marginal stability, -- encoded in the residues at the poles which
we encounter while deforming the new contour to the old one, --
is exponentially suppressed relative to the
leading contribution to the degeneracy. As a result the change in the
statistical entropy as we move across the wall of marginal stability
is exponentially suppressed. This is consistent with the fact that the
entropy of the corresponding black hole is controlled solely by the
charges and is independent of the asymptotic moduli due to
attractor mechanism. Thus as the asymptotic values of the
moduli fields move across the wall of marginal stability the
answer for the black hole entropy does not change. Our results
indicate however that if we are able to incorporate non-perturbative
(in inverse charges) 
corrections in the computation of the 
black hole entropy then we should see 
a dependence of the entropy on the asymptotic
values of the moduli fields, possibly along the line of
\cite{0010222,0101135}.

Since we use duality invariance to find how the contour should be
deformed as we move across various walls of marginal stability,
one could ask if there is any non-trivial test of duality that one
could perform. If we
could identify duality transformations which leave a region
invariant, -- in a sense that will be made precise in
\S\ref{sdualtwo}, --
then the expression for the degeneracy should not change
under such a duality transformation. This requires that
 under such a duality transformation either the
contour should remain unchanged or it should move to a new
position such that in deforming it from the new to the old
position we do not encounter a pole in the integrand. 
This can then be explicitly tested.
There is also
a possibility that a duality transformation 
maps one region to another such that the degeneracy
in each of these regions can be computed directly. In this case 
we have an independent result of how the contour should change
as we move from the first region to the second and this can then be
compared with the predictions coming from duality. 
We identify some duality transformations
of these types and carry out the required consistency checks.

The rest of the paper is organized as follows. In \S\ref{sreview}
we review the results of
\cite{0605210,0607155,0609109}
about the degeneracy of quarter BPS
dyons in a class of $\NN=4$ supersymmetric string theories.
In \S\ref{smarginal} we determine the locations of the
walls of marginal stability
in $\NN=4$ supersymmetric string theories. In \S\ref{sdualtwo}
we determine how the different domains of the moduli space,
bounded by the marginal stability walls, are mapped to each other
under T- and S-duality symmetries of the theory and
use this information to determine how the spectrum should change as
we pass through a particular marginal stability wall. In \S\ref{sblack}
we show that the change in the statistical entropy as we
move across the marginal stability walls is non-leading compared to
the full entropy. In \S\ref{stest} we perform some tests of
S-duality invariance of the theory.

Finally we would like to remind the reader 
that our analysis will focus on the
marginal stability walls associated with decay of the dyon into a
pair of half-BPS states. It will be worth exploring if there are
interesting phenomena associated with decay into a pair of
quarter BPS states.

Some related issues have been discussed in \cite{appear}.

\sectiono{Dyon Spectrum
in a Class of $\NN=4$ Supersymmetric Models} \label{sreview}

In this section we shall review the dyon spectrum in a
class of $\NN=4$ supersymmetric
models analyzed in \cite{0510147,0602254,0605210,
0607155,0609109}. Somewhat different approaches leading to
similar results have been developed in 
\cite{9607026,0505094,0506249,0603066,0612011}.

These theories are constructed by taking a $\ZZZ_N$ orbifold of
type IIB string theory on $\MM\times S^1\times \wt S^1$ where
$\MM$ is either K3 or $T^4$. 
The generator $g$ of the $\ZZZ_N$ group involves
$1/N$ unit of shift along the circle $S^1$ 
together with an order
$N$ transformation
$\wt g$ in $\MM$. $\wt g$ is chosen so that it commutes with
an $\NN=4$ supersymmetry algebra of the parent theory.
Thus the final
theory has  $\NN=4$ supersymmetry. 

The description of the theory given above will be referred to as
the first
description of the theory. Another useful description is
obtained by a series of duality transformations. We first
make an S-duality transformation in the type IIB theory.  Next
we make 
an $R\to 1/R$ duality on the circle $\wt S^1$ that takes type IIB
string theory on $\MM\times S^1\times\wt S^1$ to type IIA  string
$\MM\times S^1\times\wh S^1$ where $\wh S^1$ is
the circle dual to $\wt S^1$. Finally  using 
six dimensional string-string
duality we relate this to a heterotic string theory on $T^4\times S^1
\times \wh S^1$ for 
$\MM=K3$
and type IIA string theory on
$T^4\times S^1
\times \wh S^1$ for $\MM=T^4$. 
Under this duality the transformation
$\wt g$ gets mapped to a transformation $\wh g$ that acts only as
a shift on the right-moving degrees of freedom on the world-sheet
and as a shift plus rotation on the left-moving degrees of freedom.
In the final theory, obtained by taking the orbifold of heterotic or
type IIA
string theory on $T^4\times S^1\times \wh S^1$ by a $1/N$
unit of shift along $S^1$ together with the transformation $\wh g$, all
the space-time supersymmetries come from the right-moving sector
of the world-sheet. We shall call this the second description of the
theory. 

These theories typically have several moduli fields.
In the second description one complex modulus scalar 
arises
from the axion ($a$) - dilaton ($\phi$) 
combination, where the axion by definition
is the scalar field obtained by dualizing the NSNS sector
2-form field. We shall denote by $\tau$ the combination
$a + i S$ where $S=e^{-2\phi}$. Other
moduli fields arise from the $\wh g$ invariant
components of the metric, antisymmetric tensor field and gauge
fields (in case of heterotic string theory)
along the six dimensional internal torus and 
may be encoded in an $r\times r$ matrix valued field $M$ satisfying
\be \label{eag1}
MLM^T=M, \quad M^T=M \, ,
\ee
where $L$ is a matrix with 6 eigenvalues +1 and 
$(r-6)$ eigenvalues $-1$. Here $r$ is the rank of the gauge group
and depends on the specific model being considered.
These U(1)
gauge fields  arise from the $\wh g$ invariant ten dimensional gauge fields 
(in case of heterotic string theory) as well as $\wh g$ invariant
components of the
metric and the 2-form field with one leg along the
internal   torus and one leg along the non-compact directions.

Following the chain of dualities relating the first description to the
second description one can work out the origin of the various
fields in the first description. In particular the modulus $\tau=a+iS$
can be shown to correspond to the complex structure modulus of
the torus spanned by the $S^1$ and $\wt S^1$ directions.

We shall now give the precise relationship between 
some of the moduli fields and the
geometric quantities associated with the compactification in
the second description and also
give precise expression for some of the charges
in terms of physical quantum numbers
carried by a state. To do this we need to fix a normalization convention
for the coordinates along $S^1$ and $\wh S^1$. We choose 
$x^4$ and $x^5$ to be the coordinates along $\wh S^1$
and $S^1$ respectively and choose their periods before
orbifolding to be $2\pi\sqrt{\alpha'}$ 
and $2\pi N\sqrt{\alpha'}$ respectively. Thus after
orbifolding both can be regarded as having period 
$2\pi\sqrt{\alpha'}$; however
since periodicity under a $2\pi\sqrt{\alpha'}$ translation 
action $S^1$ also
involves an action of the generator $\wh g$ of the internal $\ZZZ_N$
symmetry group, the momentum along $S^1$, measured in units
of $1/\sqrt{\alpha'}$, could be 
fractional,
-- some multiple of $1/N$, -- if
$\wh g$ acts non-trivially on the state. 
A string will be said to carry one 
unit of winding
along $S^1$ (or $\wt S^1$) if, as we go once
around the string, the $x^4$ ($x^5$) coordinate 
shifts by $2\pi\sqrt{\alpha'}$.
Thus an untwisted sector state whose coordinate along $S^1$
changes by multiples of $2\pi N\sqrt{\alpha'}$ 
will carry winding charge along
$S^1$ in
multiples of $N$, but twisted sector states can carry generic integer
winding charges.
A single H-monopole associated with $S^1$
will correspond to an
array of NS 5-branes wrapped on $\wh S^1\times T^4$
and placed at
intervals of $2\pi\sqrt{\alpha'}$ along $S^1$. Finally the original 
Kaluza-Klein monopole associated  with $S^1$,
represented by a
Taub-NUT
space with an asymptotic circle of radius $2\pi N\sqrt{\alpha'}$ 
along $x^4$,
will develop a $\ZZZ_N$ singularity at its centre after the
orbifolding and has to be 
regarded as carrying $N$ units of  Kaluza-Klein
monopole charge associated with $S^1$. 
Thus the Kaluza-Klein monopole charge associated with $S^1$
will be quantized
in units of $N$. Similar definition can be given for the $H$ and
Kaluza-Klein monopole charges associated with $\wh S^1$, but
in this case the normalization is straightforward and both the
charges are allowed to take arbitrary integer values.

Let  $x^\mu$ ($0\le
\mu\le 3$) denote the coordinates along
the non-compact coordinates.
For our analysis it will be 
useful to study in detail a subsector
of the theory in which we include only those gauge fields 
which are associated with the $4\mu$ and $5\mu$
components of the metric and the
anti-symmetric tensor field, 
only those components of $M$ which encode information
about the
components of the metric and the
anti-symmetric tensor field
along $S^1\times \wh S^1$, the axion-dilaton field,
and the four dimensional metric. In this subsector there are
four gauge fields $A^{(i)}_\mu$ ($1\le i\le 4$) and 
a $4\times 4$ matrix valued field $M$ satisfying
\be\label{enewm}
M^T = M, \quad M L M^T = L, \quad L\equiv\pmatrix{0 & I_2\cr
I_2 & 0}\, .
\ee
The fields $A^{(i)}_\mu$ and $M$
are related to the ten dimensional string metric $G_{MN}$ and
2-form field $B_{MN}$ via the relations\cite{9207016,9402002}:
\ben \label{etenfour}
&& \wh G_{mn} \equiv G^{(10)}_{mn}, \quad 
\wh B_{mn} \equiv B^{(10)}_{mn}\, ,  
\qquad 
m,n=4,5\, , 
\nonumber  \\ 
&& M =  
\pmatrix{ \wh G^{-1} & \wh G^{-1} B \cr -\wh B \wh G^{-1} & \wh 
G - \wh B \wh G^{-1} \wh B} \nonumber \\
&& A^{(m-3)}_\mu = {1\over 2} (\wh G^{-1})^{mn} 
G^{(10)}_{m\mu} , \quad
A^{(m-1)}_\mu = {1\over 2} B^{(10)}_{m\mu} - 
\wh B_{mn} A^{(m-3)}_\mu, \nonumber \\
&& \qquad  4\le m,n\le 5, \quad 
0\le \mu, \nu \le 3 \, . 
\een

A general
dyonic state in the theory is characterized by an $r$ dimensional
electric charge
vector $\vec Q$ and an $r$ dimensional 
magnetic charge vector $\vec P$. 
However if we consider a dyon that is charged only under the
gauge fields $A_\mu^{(i)}$ ($1\le i\le 4$) introduced in
\refb{etenfour}, the corresponding charge vectors can be taken to be
four dimensional. 
If  we consider a state with 
momentum $\wh n$ and
winding $-\wh w$ along $\wh S^1$, 
momentum $n'$ and
winding $-w'$ along $S^1$,
Kaluza-Klein monopole charge  $\wh N$ 
and 
H-monopole
charge  $-\wh W$ associated with 
$\wh S^1$ and 
Kaluza-Klein monopole charge $N'$ and
H-monopole
charge $-W'$ associated with 
$S^1$, then we define the four dimensional electric
charge vector $\vec Q$ and magnetic charge vector $\vec P$
characterizing the state to be
\be\label{e2dcharge}
Q=\pmatrix{\wh n\cr n'\cr \wh w\cr w'}, \qquad 
P = \pmatrix{\wh W\cr W'\cr
\wh N\cr N'}\, .
\ee
{}From our earlier discussion it follows that $n'$ is quantized in
units of $1/N$, $N'$ is quantized in units of $N$ and all other
quantum numbers appearing in \refb{e2dcharge} are quantized
in integer units. The precise relation between $Q$, $P$ and the
electric and magnetic charges associated with the gauge fields
$A_\mu^{(i)}$ has been derived 
 in \cite{0508042} by working
in the
$\alpha'=16$ unit.

Restricted to this subspace the 
T-duality transformation of the theory is parametrized by
a $4\times 4$ matrix $\Omega$ preserving the charge lattice
and satisfying
\be \label{eag9}
\Omega^T L \Omega=L\, .
\ee 
The action of
T-duality takes us from a charge vector $(\vec Q, \vec P)$ to another
charge vector $(\vec Q',\vec P')$ given by
\be \label{sadd1}
\vec P'= (\Omega^T)^{-1}  \, \vec P, \qquad \vec Q'=(\Omega^T)^{-1} 
\, \vec Q\, ,
\ee
and the moduli field $M$ to $M'$ given by
\be \label{emtrs}
M'=\Omega M \Omega^T \, ,
\ee
leaving $\tau$ unchanged.
{}From eqs.\refb{eag9}, \refb{sadd1} it follows that the following
inner products are invariant under T-duality transformation
\ben\label{echsq}
&& Q^2 \equiv Q^T L Q
= 2(\wh n \wh w + n'w'), \qquad P^2 \equiv P^TLP
= 2(\wh N \wh W
+N'W'), \nonumber \\ &&
P\cdot Q \equiv P^TLQ
= \wh N \wh n + \wh W \wh w+ N' n' + W' w'\, .
\een
These combinations are independent of the moduli fields $M$.
Using the moduli field we can construct more general T-duality
invariant combinations like $Q^TMQ$, $P^TMP$ and $Q^TMP$.
We shall make use of these quantities in \S\ref{smarginal}.

In the full theory $\vec Q$ and $\vec P$ are $r$ dimensional vectors
and $\Omega$ and $M$ are $r\times r$ matrices, and $L$ is also an
$r\times r$ matrix with 6 eigenvalues 1 and $(r-6)$ eigenvalues $-1$.
However the form of eqs.\refb{eag9}-\refb{emtrs} 
remains
the same.

The theories under consideration also have  
S-duality symmetry which leaves the field $M$ unchanged,
changes the vector $(\vec Q, \vec P)$
to another vector $(\vec Q'', \vec P'')$ via the formula
\be \label{sadd3}
\pmatrix{\vec Q''\cr \vec P''} = \pmatrix{\alpha & \beta\cr \gamma
 & \delta}
\pmatrix{\vec Q\cr \vec P}\, , \qquad 
\pmatrix{\alpha & \beta\cr \gamma
 & \delta}\in \Gamma_1(N)\, ,
\ee
and transforms $\tau=a+iS$ to
\be \label{sadd4}
\tau''={\alpha\tau+\beta\over \gamma\tau+\delta}\, ,
\ee
where the $\Gamma_1(N)$ group of matrices 
is defined by the conditions
\be \label{egammacond}
\alpha\delta - \beta\gamma=1, \qquad \alpha, \delta = 
\hbox{1 mod $N$}, \qquad \gamma = \hbox{0 mod $N$},
\qquad \beta\in \ZZZ\, .
\ee

We now consider in the first description of this theory
 a configuration with a single D5-brane wrapped on 
 $\MM\times S^1$, $Q_1$ D1-branes wrapped on $S^1$, a single
 Kaluza-Klein monopole associated with the circle $\wt S^1$,
 momentum $-n/N$ along $S^1$ and momentum $J$
 along $\wt S^1$\cite{0505094}. Since a 
 D5-brane wrapped on $\MM$
 carries, besides the D5-brane charge, $-\zeta$ 
 units of D1-brane charge
 with $\zeta$ given by the Euler character of $\MM$ 
  divided by 24 \cite{9511222}, the net 
  D1-brane charge carried by the system
  is $Q_1-\zeta$.
 By following the chain of dualities described earlier and a
 suitable sign convention for the charges in the first description,
 we can map this to a configuration
 in the second description with momentum $-n/N$ 
 along $S^1$, a single Kaluza-Klein monopole associated with 
 $\wh S^1$,
  $(-Q_1+\zeta)$  NS 5-brane charge wrapped along $T^4\times
  S^1$, $J$  NS 5-brane charge wrapped along 
  $T^4\times \wh S^1$
  and  unit fundamental string winding charge along 
  $S^1$\cite{0605210}. In particular the Kaluza-Klein monopole
  charge associated with $\wt S^1$ in the first description gets mapped to
  the fundamental string winding number along $S^1$ in the second
  description and the D5-brane wrapped on $\MM\times S^1$ in the
  first description gets mapped to Kaluza-Klein monopole charge
  associated with $\wh S^1$ in the second description.
  Using \refb{e2dcharge} we see that 
  this corresponds to the charge vectors\footnote{Recall that an
  NS 5-brane wrapped on $T^4\times S^1$ represents an H-monopole
  associated with $\wh S^1$ and an
  NS 5-brane wrapped on $T^4\times \wh S^1$ 
  represents an H-monopole
  associated with $S^1$.}
  \be \label{echvec}
  Q=\pmatrix{0 \cr -n/N \cr 0 \cr -1}, \quad P = \pmatrix{Q_1-\zeta
 \cr -J\cr 1 \cr 0}\, .
  \ee
    This gives
 \be\label{eqdef}
 Q^2  
 = 2 n/N, \qquad P^2   
 = 2 (Q_1- \zeta   ), \qquad Q\cdot P = J\, .
\ee

The counting of states of this system in the weak coupling region
of the first description was carried out in 
\cite{0605210,0607155,0609109}. 
We shall now summarize the results of this analysis.
We denote by 
$d(\vec Q,\vec P)$ the number of bosonic minus fermionic
quarter BPS supermultiplets carrying a given set of charges
$(\vec Q, \vec P)$, a supermultiplet being considered bosonic
(fermionic) if it is obtained by tensoring the basic 64 dimensional
quarter BPS supermultiplet, with helicity ranging from $-{3\over 2}$
to ${3\over 2}$,  with a supersymmetry singlet  
bosonic (fermionic) state.
Our result for $d(\vec Q,\vec P)$  is
\be\label{egg1int}
d(\vec Q,\vec P) = {1\over N}\, \int _\CC d\wt\rho \, 
d\wt\sigma \,
d\wt v \, e^{-\pi i ( N\wt \rho Q^2
+ \wt \sigma P^2/N +2\wt v Q\cdot P)}\, {1
\over \wt\Phi(\wt \rho,\wt \sigma, \wt v)}\, ,
\ee
where $\CC$ is a three real dimensional subspace of the
three complex dimensional space labelled by $(\wt\rho=\wrh_1+i
\wrh_2,\ws=\ws_1+i\ws_2,\wv=\wv_1+i\wv_2)$. $\CC$ corresponds
to the subspace:\footnote{The sign of $\wt v_2$ chosen here differs
from the ones used in \cite{0605210,0607155,0609109}. 
As will be discussed later, this choice of sign is valid for a
specific choice of sign of the axion field; for the other choice
the sign of $\wv_2$ needs to be reversed. The other difference is
that we have put the $M_3<<M_1, M_2$ condition to make
explicit the fact that while interpreting $d(\vec Q, \vec P)$ as the
coefficients of the Fourier expansion of $1/\wt\Phi(\wrh,\ws,\wv)$
as in eqs.\refb{efo1}, \refb{efo2} 
we need to first carry out the expansion in
powers of $e^{2\pi i\wrh}$ and $e^{2\pi i\ws/N}$ and then carry out
the expansion in powers of $e^{-2\pi i\wt v}$. This was implicit
in the results of \cite{0605210,0607155,0609109} but was not
stated explicitly.}
\bea{ep2kk}
 \wt \rho_2=M_1, \quad  \wt\sigma_2 = M_2, \quad
 \wt v_2 = -M_3, \nonumber \\
 0\le  \wt\rho_1\le 1, \quad
0\le  \wt\sigma_1\le N, \quad 0\le  \wt v_1\le 1\, ,
\een
$M_1$, $M_2$ and $M_3$ being large but fixed positive
numbers with $M_3<< M_1, M_2$, and
\bea{edefwtphi}
&& \wt \Phi(\wt \rho,\wt \sigma,\wt v ) =
e^{2\pi i (\wt \alpha\wt\rho + \wt \gamma\ws 
+ \wt v)} \nonumber \\
&& \qquad \times \prod_{b=0}^1\, 
 \prod_{r=0}^{N-1}
\prod_{k'\in \zzz+{r\over N},l\in\zzz,j\in 2\zzz+b
\atop k',l\ge 0, j<0 \, {\rm for}
\, k'=l=0}
\left( 1 - \exp\left(2\pi i ( k'\wt \sigma   +  l\wt \rho +  j\wt v)
\right)\right)^{
\sum_{s=0}^{N-1} e^{-2\pi i sl/N } c^{(r,s)}_b(4k'l - j^2)} \, .
\nonumber \\
\eea
The coefficients $c^{(r,s)}_b(u)$, $\wt\alpha$,
$\wt\gamma$ encode information about
the spectrum of two dimensional superconformal $\sigma$-model
with target space $\MM$ and are defined as follows.
First we define
\be\label{esi4aint}
F^{(r,s)}(\tau,z) \equiv {1\over N} Tr_{RR;\wt g^r} \left(\wt g^s
(-1)^{F_L+F_R}
e^{2\pi i \tau L_0} 
e^{-2\pi i \bar\tau \bar L_0}
e^{2\pi i F_L z}\right), \qquad 0\le r,s\le N-1\, ,
 \ee
where
$Tr_{RR;\wt g^r}$ denotes trace 
over all the Ramond-Ramond (RR) sector 
states twisted by $\wt g^r$ in the SCFT described above, 
$L_n$, $\bar L_n$ denote the left- and right-moving
Virasoro generators and $F_L$ and $F_R$ 
denote the world-sheet fermion 
numbers
associated with left and right-moving sectors in this 
SCFT. In defining $L_0$ and $\bar L_0$ of a state we subtract 
$c_L/24$ and $c_R/24$ from the conformal weights of the
corresponding operators so that the RR sector ground state has
$L_0=\bar L_0=0$.
Due to the insertion of $(-1)^{F_R}$ factor in the trace the
contribution to $F^{(r,s)}$ comes only from the $\bar L_0=0$
states. As a result $F^{(r,s)}$ does not depend on $\bar\tau$.
Furthermore, using the existence of an $SU(2)_L\times SU(2)_R$
R-symmetry current algebra in this theory
one can show that $F^{(r,s)}(\tau,z)$ have expansions of the form
\be\label{enewint}
F^{(r,s)}(\tau,z) =\sum_{b=0}^1\sum_{j\in2\zzz+b, n\in \zzz/N
\atop 4n-j^2\ge -b^2} 
c^{(r,s)}_b(4n -j^2)
e^{2\pi i n\tau + 2\pi i jz}\, ,
\ee
for some coefficients $c^{(r,s)}_b(u)$.  This defines the coefficients
$c^{(r,s)}_b(u)$.
We now define
\be\label{eqrsrev}
Q_{r,s} = N\, 
\left( c^{(r,s)}_0(0)+ 2 \, c^{(r,s)}_1(-1)\right)\, ,
\ee
\be\label{enn9d}
\wt \alpha={1\over 24N} \, Q_{0,0} - {1\over 2N}
\, \sum_{s=1}^{N-1} Q_{0,s}\, {e^{-2\pi i s/N}\over
(1-e^{-2\pi i s/N})^2 } \, 
, \qquad 
\wt \gamma= {1\over 24N} \, Q_{0,0}  \, .
\ee
This defines all the coefficients appearing in \refb{egg1int}.

As an alternative to \refb{egg1int}, \refb{ep2kk}
we can express $d(\vec Q,\vec P)$ as
\be\label{efo1}
d(\vec Q, \vec P) = g\left({N\over 2} Q^2 , {1\over 2\, N}\, P^2,
Q\cdot P\right)\, ,
\ee
where $g(m,n,p)$ are
the coefficients of Fourier expansion of the function
$1/ \wt\Phi(\wt \rho,\wt \sigma, \wt v)$:
\be\label{efo2}
{1
\over \wt\Phi(\wt \rho,\wt \sigma, \wt v)}
=\sum_{m,n,p} g(m,n,p) \, e^{2\pi i (m\, \wt \rho + n\,
\wt\sigma
+ p\, \wt v)}\, .
\ee

Let us denote by $G$ the group of $4\times 4$ matrices generated by:
\ben \label{egroup}
   \pmatrix{A&B\cr C&D}  &=& \pmatrix{ a & 0 & b & 0 \cr
     0 & 1 & 0 & 0\cr c & 0 & d & 0\cr 0 & 0 & 0 & 1}\, ,
   \qquad ad-bc=1, \quad \hbox{$c=0$ mod $N$, \quad $a,d=1$
     mod $N$}
   \nonumber \\
   \pmatrix{A&B\cr C&D}  &=& 
   \pmatrix{0 & 1 & 0 & 0 \cr -1 & 0 & 0 & 0\cr
     0 & 0 & 0 & 1\cr 0 & 0 & -1 & 0}\, , \nonumber \\
   \pmatrix{A&B\cr C&D}  &=& 
   \pmatrix{ 1 & 0 & 0 & \mu \cr
     \lambda & 1 & \mu & 0\cr 0 & 0 & 1 & -\lambda\cr
     0 & 0 & 0 & 1}\, , \qquad \lambda, \mu \in \ZZZ,
\een
and let $\wt G$ denote the group of $4\times 4$ matrices
satisfying the requirement that:
\be\label{ebelong}
\wt g\in \wt G \quad \hbox{iff} \quad U^{-1} \, \wt g U\in G\, ,
\ee
where\footnote{Even though $U$ contains factors of $\sqrt{N}$
the elements of $\wt G$ are integers due to the
fact that for $\pmatrix{A & B\cr C & D}\in G$, the elements of
$C$ are multiples of $N$\cite{0510147}.}
\be\label{edefu}
U = \pmatrix{0 & 0 & 0 & -1/\sqrt N\cr -\sqrt N & 0 & 0 & 0\cr
0 & \sqrt N & 0 & 0\cr 0 & 0 & -1/\sqrt N & 0}\, .
\ee
An element $\pmatrix{\wt A & \wt B\cr \wt C & \wt D}\in \wt G$
induces a transformation on $(\wrh,\ws,\wv)$ via the relations:
\be\label{enatural}
\pmatrix{\wrh' & \wv'\cr \wv' & \ws'}
= (\wt A \wt\Omega +\wt B) (\wt C\wt\Omega + \wt D)^{-1},
\qquad \wt\Omega\equiv \pmatrix{\wrh & \wv\cr \wv & \ws}\, ,
\ee
and one can show that
$\wt\Phi$ transforms as a Siegel modular form of weight $k$
under this transformation\cite{0609109}:
\be \label{ewtphitrs}
\wt\Phi(\wrh',\ws',\wv') 
= \det(\wt C\wt\Omega+\wt D)^k \, \wt\Phi(\wrh,\ws,\wv) \, ,
\ee
where
\be\label{ekvalue}
k={1\over 2}\, \sum_{s=0}^{N-1} \, c_0^{(0,s)}(0)\, .
\ee

This finishes our review of the main results. However
two points about the degeneracy formula given  above need 
special mention. Eqs.\refb{egg1int} and \refb{efo1} are equivalent
only if the sum over $m$, $n$, $p$ in \refb{efo2}
are convergent on the contour $\CC$. For the choice of $\CC$
given in \refb{ep2kk} this requires that $\wt\Phi(\wrh,\ws,\wv)^{-1}$
has a power series expansion in positive powers of $e^{2\pi i\wrh}$
and $e^{2\pi i\ws/N}$ except possibly for a finite number of negative
powers, and that the  coefficient  of any
given term in this double power series expansion has an expansion
in positive powers of $e^{-2\pi i \wv}$ 
except possibly for a finite number of negative
powers.
This in particular 
requires that the sum over $m$ and $n$ in \refb{efo2}
are bounded from below,
and that for fixed $m$ and $n$
the sum over $p$ is bounded from above.  By examining the
formula \refb{edefwtphi} for $\wt\Phi$ and the result that the
coefficients $c^{(r,s)}_b(u)$ are non-zero only for $u\ge -b^2$,
we can verify that with the exception of the contribution from the
$k'=l=0$ term in this product, the other terms, when expanded
in a power series expansion in $e^{2\pi i\wrh}$, $e^{2\pi i\ws}$
and $e^{2\pi i\wv}$, 
does have the form of \refb{efo2} with $p$ bounded from above
(and below)
for fixed $m$, $n$.
However for the $k'=l=0$ term, which  gives a contribution
$e^{-2\pi i \wv} / (1 - e^{-2\pi i \wv})^2$, 
there is an ambiguity in
carrying out the series expansion. We could either use the form given
above and expand the denominator in a series 
expansion in $e^{-2\pi i \wv}$ so that the criterion described
above is satisfied, or express it as
$e^{2\pi i \wv} / (1 - e^{2\pi i \wv})^2$ and expand it in 
a series 
expansion in $e^{2\pi i \wv}$ in which case the sum over $p$
will be bounded from below rather than from
above.  One finds that
depending on the angle between $S^1$ and $\wt S^1$,
only one of these expansions produces the degeneracy
correctly via \refb{efo2}\cite{0605210}. The physical 
spectrum actually changes as this angle
passes through $90^\circ$ since at this point the system is only
marginally stable.
On the other hand our degeneracy formula
\refb{egg1int}, \refb{ep2kk} 
implicitly assumes that we have
expanded this factor in powers
of $e^{-2\pi i \wv}$ since only in this case the sum over $p$ in
\refb{efo2} is bounded from above for fixed $m$, $n$.
Thus as it stands the expression for $d(\vec Q,\vec P)$
given in \refb{egg1int}, \refb{ep2kk} 
is valid for a
specific range of values of the angle between $S^1$ and $\wt S^1$,
which, in the second
description of the system, corresponds to the sign of the axion field.
For the other sign of the axion we need to carry out the integral
over a different contour $\wh \CC$
defined as
\ben \label{ecchat}
 \wt \rho_2=M_1, \quad  \wt\sigma_2 = M_2, \quad
 \wt v_2 = M_3, \nonumber \\
 0\le  \wt\rho_1\le 1, \quad
0\le  \wt\sigma_1\le N, \quad 0\le  \wt v_1\le 1\, ,
 \een
where $M_1$, $M_2$ and $M_3$ are large positive numbers
with $M_3<<M_1,M_2$.
In the
convention that we shall be using in this paper, we need to use
the contour $\CC$ for positive sign of the axion and the
contour $\wh \CC$ for negative sign of the axion.

It turns out that walls of marginal stability, -- codimension
one subspaces of the asymptotic moduli space on which
the BPS mass of the 
system becomes equal to the sum of masses of two  or more
other
states carrying the same total charge, -- are quite generic for
quarter BPS states in $\NN=4$ supersymmetric string
theories\cite{9712211}, and we expect the 
spectrum to change discontinuously as
the asymptotic moduli fields pass through any of the
walls of marginal 
stability.\footnote{If a state becomes
marginally stable on a surface of codimension $\geq 2$, then we can
always move around this subspace in going from one point to 
another and hence the spectrum cannot change discontinuously.}
Thus the expression for the degeneracy given in this section 
holds only in
a finite region of the moduli space, bounded by the walls of 
marginal stability. This will be discussed in more detail in
\S\ref{smarginal}.

Another point about the formula \refb{egg1int} is that although 
it was
derived for special
charge vectors $\vec Q$, $\vec P$ described in \refb{echvec}, it has
been expressed as a function of the T-duality invariant combinations
$P^2$, $Q^2$ and $Q\cdot P$.
Even though we expect T-duality to be  a symmetry of the theory,
it is not guaranteed that the formula written in this fashion hold
for all charge vectors. First of all a T-duality transformation acts not
only on the charges but also on the asymptotic moduli. 
Had the spectrum been
independent of the asymptotic moduli, we could have demanded
that the spectrum remains invariant under T-duality transformation
of the charges. However if 
a T-duality
transformation takes the asymptotic moduli fields 
across a wall of marginal stability, then
all we can say is that the spectrum remains unchanged under a
simultaneous T-duality transformation of the moduli fields
and the charges, but if we are sitting at a fixed point in the moduli
space then the spectrum is not invariant under T-duality
transformation on the charges. Second point is that even if we
ignore the issues related to the walls of marginal stability,
two charge vectors carrying the same values of $P^2$, $Q^2$
and $Q\cdot P$ may not necessarily
be related by a T-duality
transformation.\footnote{Generically they are related by a
continuous T-duality transformation but only a discrete subgroup
of this is a genuine symmetry of the theory.} In that case the
degeneracy of states for these two charge vectors could be
different. An example of this is that a state that carries
fractional momentum along
$S^1$ can never be related to a state carrying
integer momentum along $S^1$, although they may carry 
same values of $Q^2$, $P^2$ and $Q\cdot P$. Both these issues,
together with the S-duality transformation properties of the
degeneracy formula, will be discussed in \S\ref{sdualtwo}.

\sectiono{Walls of Marginal Stability}
\label{smarginal}

As has been briefly mentioned in \S\ref{sreview}, the 
degeneracy formula given in \refb{egg1int},  \refb{ep2kk}
is expected to be valid within a certain region of the
moduli space bounded by codimension one subspaces on which the
BPS state under consideration becomes marginally stable. As we
cross this subspace of the moduli space, the spectrum can change
discontinuously. In this section we 
shall study in some detail the locations
of these walls of marginal stability so that we can identify the region
within which our degeneracy formula will remain valid.

Let us 
consider a state carrying electric charge $\vec Q$ and magnetic charge
$\vec P$ and examine under what condition it can decay into a
pair of half-BPS states.
This happens
when its mass is equal to the sum of the masses of a pair of half BPS
states whose electric and magnetic charges add up to $\vec Q$ and
$\vec P$ respectively. Since for half BPS states the electric and magnetic
charges must be parallel, these pair of states must
have charge vectors of
the form $(a\vec M, c \vec M)$ and $(b \vec N, d \vec N)$ for some
constants $a$, $b$, $c$, $d$ and a pair of $r$-dimensional vectors
$\vec M$, $\vec N$. We shall normalize $\vec M$, 
$\vec N$ such that 
\be \label{enorm}
ad-bc=1\, .
\ee
Then the requirement that the charges add up to $(\vec Q, \vec P)$ gives
\be \label{esm2}
\vec M = d\vec Q - b\vec P, \qquad \vec N = -c\vec Q + a\vec P\, .
\ee
Thus the charges of the decay products are given by
\be \label{escale0}
(ad\vec Q-ab\vec P, cd\vec Q-cb\vec P) \quad \hbox{and} \quad
(-bc\vec Q + ab\vec P, -cd\vec Q+ad\vec P)\, .
\ee
Note  that under the scale
transformation
\be \label{escale1}
\pmatrix{a & b\cr c & d}\to 
\pmatrix{a & b\cr c & d} \, \pmatrix{\lambda & 0\cr 0 
& \lambda^{-1}}
\ee
eqs.\refb{enorm} and \refb{escale0} remain unchanged.
There is another discrete transformation
\be \label{esc12}
\pmatrix{a & b\cr c & d}\to 
\pmatrix{a & b\cr c & d} \, \pmatrix{0 & 1\cr -1 & 0}\, ,
\ee
which  leaves \refb{enorm} unchanged and exchanges the
two decay products in \refb{escale0}.
A pair of matrices $\pmatrix{a & b\cr c & d}$ related by
\refb{escale1} or \refb{esc12} describe identical decay channels.

In order that the charge vectors of the decay products
given in \refb{escale0} satisfy the
charge quantization rules we must ensure that $a\vec M
=ad\vec Q-ab\vec P$
and $b\vec N=-bc\vec Q + ab\vec P$ belong to the lattice of electric
charges and that $c\vec M=cd\vec Q-cb\vec P$ and 
$d\vec N=-cd\vec Q+ad\vec P$ belong to
the lattice of magnetic charges. 
For the charge vectors $\vec Q$, $\vec P$ given in
\refb{echvec} this would require 
\be \label{egen}
ad,ab,bc\in \ZZZ, \qquad 
cd\in N\ZZZ\, .
\ee
The condition $cd\in N\ZZZ$ comes from the requirement
that $cd\vec Q-cb\vec P$ is an allowed magnetic charge.
In particular for a $\vec Q$ of the form \refb{echvec}, a magnetic
charge $cd\vec Q$ represents a state with
Kaluza-Klein monopole charge $-cd$
associated with $S^1$. Since this charge is quantized in units
of $N$, $cd$ must be a multiple of $N$.
We shall denote by $\AAA$ the set of matrices $\pmatrix{a & b\cr c
& d}$ subject to the equivalence relations \refb{escale1}, \refb{esc12}
and satisfying \refb{enorm}, \refb{egen}.

It is instructive to determine the structure of the set $\AAA$. We shall
first
show that using the scale transformation \refb{escale1} we can
always choose $a$, $b$, $c$ and $d$ to be integers and furthermore
the solution is unique for given $ad$, $ab$, $bc$ and $cd$. 
Since
$ad$, $ab$, $bc$ and $cd$ are all integers, we can express them as
products of prime factors:
\be \label{eprime1}
|ab| = \prod_i \, p_i^{r_i}, \qquad |cd| = \prod_i p_i^{s_i}, \qquad
|ad| = \prod_i p_i^{u_i}, \qquad |bc| = \prod_i p_i^{v_i},
\ee
where the product over $i$ runs over the prime numbers $p_i$ and
$r_i$, $s_i$, $u_i$ and $v_i$ are non-negative integers satisfying
\be \label{eprime2}
r_i + s_i = u_i + v_i \quad \forall\, i\, .
\ee
Furthermore since $ad$ and $bc$ differ by 1, they cannot have a
common factor. This shows that either $u_i$ or $v_i$ must vanish:
\be \label{eprime3}
u_i \, v_i = 0 \quad \forall \, i\, .
\ee
Let us now look for integer $a$, $b$, $c$ and $d$ satisfying
\refb{eprime1}. For this we use the ansatz:
\be \label{eprime4}
|a| = \prod_i \, p_i^{a_i}, \qquad |b| = \prod_i p_i^{b_i}, \qquad
|c| = \prod_i p_i^{c_i}, \qquad |d| = \prod_i p_i^{d_i},
\ee
where $a_i$, $b_i$, $c_i$ and $d_i$ are non-negative integers.
Eq.\refb{eprime1} now gives
\be \label{eprime5}
a_i + b_i = r_i, \qquad c_i+d_i = s_i, \qquad a_i+d_i=u_i, \qquad
b_i + c_i = v_i\, .
\ee
Now \refb{eprime3} tells us 
that for any given $i$ either $u_i$ or $v_i$ (or both) are
zero. If $u_i=0$ then the only possible solution to
\refb{eprime5} is
\be \label{eprime6}
a_i=0, \qquad d_i=0, \qquad b_i=r_i, \qquad c_i=s_i\, .
\ee
On the other hand if $v_i=0$ then we must have
\be \label{eprime7}
b_i=0, \qquad c_i=0, \qquad a_i=r_i, \qquad d_i=s_i\, .
\ee
This gives a unique expression for $|a|$, $|b|$, $|c|$ and $|d|$ using
prime factorization. Up to an overall factor of $-1$ which can be
removed with the help of the scale transformation 
\refb{escale1}, we can
determine the signs of $a$, $b$, $c$ and $d$ in terms of the
signs of $ab$, $cd$, $ad$ and $bc$.

The requirement that $cd$ is a multiple of $N$ implies that the
prime factors of $N$ are shared by $c$ and $d$. In case $N$ is
prime either $c$ or $d$ must be a multiple of $N$. Using 
the freedom \refb{esc12} we can ensure that $c$ is a multiple
of $N$.  In this case the matrices $\pmatrix{a & b\cr c & d}$ 
describe elements of $\Gamma_0(N)$ 
modulo multiplication
by $-1$.\footnote{$\Gamma_0(N)$ contains matrices
$\pmatrix{a&b\cr c&d}$ satisfying $ad-bc=1$, $a,b,d\in\ZZZ$,
$c\in N\ZZZ$.}
Since $\Gamma_0(2)=\Gamma_1(2)$, for $N=2$ the set of matrices
$\pmatrix{a&b\cr c&d}$ may be identified as the elements of
$\Gamma_1(N)$ modulo multiplication by $-1$. On the other
hand using the freedom of multiplication by $-1$ we can convert
any $\Gamma_0(3)$ matrix to a $\Gamma_1(3)$ matrix.
Thus for $N=3$ the matrices $\pmatrix{a &b\cr c & d}$ may be
chosen to be $\Gamma_1(3)$ matrices.

The case $N=1$ with $\MM=K3$, corresponding to heterotic string
theory on $T^6$ in the second description, is somewhat special. In
this case the set $\AAA$ consists of $PSL(2,\ZZZ)$ matrices subject
to the equivalence relation \refb{esc12}.

We shall now determine the wall of marginal stability corresponding
to the decay channel given in \refb{escale0}.
Our starting point will be the formula for the mass $m(\vec Q,
\vec P)$
of a BPS state carrying electric charge $\vec Q$ and magnetic charge
$\vec P$\cite{9507090,9508094}
\ben \label{esm1}
m(\vec Q, \vec P)^2 &=& {1\over S_\infty} 
(Q - \bar\tau_\infty P)^T
(M_\infty + L) (Q -\tau_\infty P) \nonumber \\
&&
+ 2 \left[ (Q^T (M_\infty + L) Q) (P^T (M_\infty + L) P)
- (P^T (M_\infty + L) Q)^2\right]^{1/2}\, , \nonumber \\
\een
where $\tau=a+iS$ and the subscript $\infty$ denotes asymptotic
values of various fields. This expression is manifestly invariant under
the T- and S-duality transformations described in 
eqs.\refb{eag9}-\refb{emtrs}
and \refb{sadd3}-\refb{egammacond}.
In order that the state $(\vec Q,\vec P)$ is marginally stable against decay
into $(ad\vec Q-ab\vec P, cd\vec Q-cb\vec P)$ and $(-bc\vec Q
+ab\vec P, -cd\vec Q+ad\vec P)$, we need
\be \label{esm3}
m(\vec Q,\vec P) = m(ad\vec Q-ab\vec P, cd\vec Q-cb\vec P)
+ m(-bc\vec Q
+ab\vec P, -cd\vec Q+ad\vec P)\, .
\ee
Using \refb{esm1}, \refb{esm3} 
and some tedious algebra, we arrive at the
condition
\be \label{esm4}
\left(a_\infty - {ad+bc\over 2cd}\right)^2 
+ \left( S_\infty +{E\over 2 cd}
\right)^2 = {1\over 4 c^2 d^2} (1 + E^2)\, ,
\ee
where
\be \label{esm5}
E \equiv { cd (Q^T (M_\infty + L) Q)
+ ab (P^T (M_\infty + L) P) -(ad + bc) (P^T (M_\infty + L) Q) \over 
\left[ (Q^T (M_\infty + L) Q) (P^T (M_\infty + L) P)
- (P^T (M_\infty + L) Q)^2\right]^{1/2}
}\, .
\ee
Note that $E$ depends on $M_\infty$, the constants $a,b,c,d$ and
the charges $\vec Q$, $\vec P$, but is 
independent of $\tau_\infty$.
Thus for fixed $P$, $Q$ and $M_\infty$, the wall of marginal
stability describes a circle in the $(a_\infty, S_\infty)$
plane with radius 
\be \label{esm6}
R = \sqrt{1 + E^2}/2|cd|\, ,
\ee 
and center at
\be \label{esm7}
C = \left( {ad+bc\over 2cd}, -{E\over 2 cd}\right)\, .
\ee
This circle 
intersects the real $\tau_\infty$
axis at
\be \label{eireal}
{a/c} \quad \hbox{and} \quad {b/d}\, .
\ee

The cases where either $c$ or $d$ vanish require special attention.
First consider the case $c=0$. In this case the condition
$ad-bc=1$ implies that $a=d=1$. By taking the $c\to 0$
limit of \refb{esm4}, \refb{esm5} 
we see that  the wall of marginal
stability becomes a straight line in the $(a_\infty, S_\infty)$ plane
for a fixed $M_\infty$:
\be\label{ecurve1}
a_\infty - {b (P^T (M_\infty + L) P)-(P^T (M_\infty + L) Q)  \over 
\left[ (Q^T (M_\infty + L) Q) (P^T (M_\infty + L) P)
- (P^T (M_\infty + L) Q)^2\right]^{1/2}
}\, S_\infty -b =0\, .
\ee
On the other hand for $d=0$ we have $bc=-1$ and
we can choose $b=-1$, $c=1$. In the $d\to 0$ limit of \refb{esm4}, 
\refb{esm5} we get
another straight line
\be\label{ecurve2}
a_\infty - {  a (P^T (M_\infty + L) P)  -(P^T (M_\infty + L) Q) \over 
\left[ (Q^T (M_\infty + L) Q) (P^T (M_\infty + L) P)
- (P^T (M_\infty + L) Q)^2\right]^{1/2}
}\, S_\infty -a =0\, .
\ee
We now notice that this has exactly the same form as \refb{ecurve1}
with $b$ replaced by $a$. Thus these do not give rise to new walls
of marginal stability. In fact the $c=0$ and $d=0$ cases are related
by the equivalence relation \refb{esc12}.

In order to get some insight into the geometric structure of the domain
bounded by these marginal stability walls it will be useful to
study the possible intersection points of these walls in the upper
half $\tau_\infty$ plane. Let us consider a pair of such walls
characterized by the matrices $\pmatrix{a_1 & b_1\cr c_1 & d_1}$
and $\pmatrix{a_2 & b_2\cr c_2 & d_2}$. 
A convenient
procedure for studying their intersection is to convert one
of them (say the first one) to a straight line
by an SL(2,$\ZZZ)$ transformation.
We define
\be \label{etrs1}
\tau'_\infty \equiv a'_\infty + i S'_\infty = {d_1 \tau - b_1\over
-c_1\tau + a_1}\, .
\ee
Then it is easy to see that in the $(a'_\infty, S'_{\infty})$ plane
the two walls get mapped to the curves
\be \label{etrs2}
a'_\infty + {   \wt P^T (M_\infty + L) \wt Q  \over 
\left[ (\wt Q^T (M_\infty + L) \wt Q) (\wt P^T (M_\infty + L) \wt
P)
- (\wt P^T (M_\infty + L) \wt Q)^2\right]^{1/2}
}\, S'_\infty = 0\, ,
\ee
and
\be \label{etrs4}
\left(a'_\infty - {\wt a\wt d+\wt b\wt c\over 2\wt c\wt 
d}\right)^2 
+ \left( S'_\infty +{\wt E\over 2 \wt c\wt d}
\right)^2 = {1\over 4 \wt c^2 \wt d^2} (1 + \wt E^2)\, ,
\ee
where
\be \label{etrs6}
\vec{\wt Q} = d_1\vec Q - b_1\vec P, \qquad 
\vec{\wt P} = -c_1\vec Q +a_1 \vec P
\, ,
\ee
\be\label{etrs7}
\pmatrix{\wt a & \wt b\cr \wt c & \wt d} =
\pmatrix{d_1 & -b_1\cr -c_1 & a_1} \pmatrix{a_2 & b_2\cr
c_2 & d_2}\, ,
\ee
\be \label{etrs5}
\wt E \equiv { \wt c\wt d (\wt
Q^T (M_\infty + L) \wt Q)
+ \wt a\wt b (\wt P^T (M_\infty + L) \wt P) 
-(\wt a\wt d + \wt b\wt c) 
(\wt P^T (M_\infty + L) \wt Q) \over 
\left[ (\wt Q^T (M_\infty + L) \wt Q) (\wt P^T (M_\infty + L) \wt
P)
- (\wt P^T (M_\infty + L) \wt Q)^2\right]^{1/2}
}\, .
\ee
If either $\wt c$ or $\wt d$ vanishes \i.e.\ if either 
$a_1/c_1=a_2/c_2$
or $a_1/c_1=b_2/d_2$ then 
\refb{etrs4} reduces to a straight line of the form \refb{ecurve1}
or \refb{ecurve2} with $P$, $Q$, $b$ (or $a$)
replaced by $\wt P$, $\wt Q$, $\wt b$ (or $\wt a$)
respectively, 
and one finds that the only point of
intersection of this line with \refb{etrs2}
in the upper half $\tau'_\infty$ plane is at
$i\infty$. In the $\tau_\infty$ plane this corresponds to the
point $a_1/c_1$. If neither $\wt c$ nor $\wt d$ vanishes, then
by
eliminating $a'_\infty$ from \refb{etrs2} and \refb{etrs4}
we get
\ben \label{etrs8}
&& (S'_\infty)^2 \, { \wt
Q^T (M_\infty + L) \wt Q\, \wt P^T (M_\infty + L) \wt
P \over (\wt Q^T (M_\infty + L) \wt Q) (\wt P^T (M_\infty + L) \wt
P)
- (\wt P^T (M_\infty + L) \wt Q)^2} \nonumber \\
&& + S'_\infty \, { \wt Q^T (M_\infty + L) \wt Q + (\wt a\wt b/\wt c
\wt d)
\wt P^T (M_\infty + L) \wt
P \over \left[ (\wt Q^T (M_\infty + L) \wt Q) 
(\wt P^T (M_\infty + L) \wt
P)
- (\wt P^T (M_\infty + L) \wt Q)^2\right]^{1/2}
} \nonumber \\
&& + {(\wt a\wt d+\wt b\wt c)^2 -1\over 4 \wt c^2 \wt d^2} = 0\, .
\een
Using the conditions $\wt a\wt d-\wt b\wt c=1$ and $\wt a, 
\wt b, \wt c, \wt d \in \ZZZ$ we see that $\wt a \wt d$ and
$\wt b \wt c$ have same signs and also that 
$(\wt a\wt d+\wt b\wt c)^2 \ge 1$. As a result for $S'_\infty\ge 0$,
each term in the left hand side of \refb{etrs8} is non-negative
and the only possible solution is 
\be \label{etrs9}
\wt b=0\quad \hbox{or} \quad
\wt a=0\, , \qquad S'_\infty = 0 \, . 
\ee
Going back to the original variables using 
\refb{etrs1}, \refb{etrs7}  we 
see that these correspond to the following two cases:
\be \label{enewcases}
{b_1\over d_1}={a_2\over c_2} \quad \hbox{or} \quad
 {b_1\over d_1}={b_2\over d_2}, \qquad \tau_\infty={b_1\over d_1}
 \, .
 \ee
Collecting all the cases together we see that the
pair of circles characterized by the matrices $\pmatrix{a_1 & b_1
\cr c_1 & d_1}$ and $\pmatrix{a_2 & b_2
\cr c_2 & d_2}$ never intersect in the interior of the upper
half plane and intersect on the real axis if and only if the sets
\be \label{etrs11a}
\left\{{a_1\over c_1}, {b_1\over d_1}\right\} \quad \hbox{and}
\quad
\left\{{a_2\over c_2}, {b_2\over d_2}\right\} \, 
\ee
have an overlap.

{}From this analysis we see that
the set of marginal stability walls divides the upper half
$\tau_\infty$ plane into many domains, with each domain
bounded by a set of walls described by a set of matrices
$\left\{\pmatrix{a_i & b_i\cr c_i & d_i}\right\}$. Since the
walls meet on the real axis or at $i\infty$, the regions have
possible vertices on the real axis or at $i\infty$, 
but never in the interior of
the upper half $\tau_\infty$ plane.  We also note that
while the shape of the walls in the $\tau_\infty$ plane
depends on the charges and the
moduli $M_\infty$, their intersetion points are determined
only in terms of the associated matrices 
$\pmatrix{a_i & b_i\cr c_i & d_i}$.

How do we determine the walls 
bordering a given domain? We shall illustrate 
this  for the domain that includes the
large $S_\infty$ region and is bounded on
the left (right) by the
line \refb{ecurve1} for $b=0$ ($b=1$). Since these lines
intersect the real $\tau_\infty$ axis at $0$ and 1 respectively,
the domain is bounded from below by  set of circle segments
in the upper half plane described by \refb{esm4},  
of which the
first one begins
at 0 and ends at some point $P_1$
on the positive real axis, the second
one begins at $P_1$ and ends at some point $P_2$ to the right
of $P_1$ etc. with the final segment ending at 1. The parameters
$a,b,c,d$ for each segment
must be chosen such that given the starting point, it travels
maximum possible distance to the right; any other segment that
travels less distance will lie beneath the maximal segment and will
not be of relevance for this computation. 
Thus in order to determine these
segments we first need to study the following general question:
given a rational number $p/q\ge0$, what is the maximum
distance a wall can travel given that it starts at $p/q$?

Let us first consider the case $p/q>0$.
We choose $p$, $q$ to be relatively prime
and $p>0$, $q>0$.
Now eq.\refb{eireal}  shows that 
the circle \refb{esm4} intersects the real axis at $a/c$ and $b/d$.
Thus either $a/c$ or $b/d$ should be equal to $p/q$. Without
any loss of generality we can choose $a/c=p/q$ and hence 
$a=p$, $c=q$ up to an irrelevant overall sign.
The constraint $ad-bc=1$ now gives
\be \label{ebbyd}
{b\over d} = {p\over q} - {1\over qd}\, .
\ee
We need to choose $d$ so as to maximize the right hand side.
Thus $d$ must be chosen to be the maximum possible negative
number subject to the requirement that $cd=qd$ is a multiple of
$N$ and $b=(pd-1)/q$ is an integer. 

In the special case where $p/q=0$, we have $a/c=0$ and
hence $a=0$. The condition
$ad-bc=1$ tells us that we have (up to an overall sign) 
$b=-1$, $c=1$. The second intersection point $b/d=-1/d$ is
maximized for maximum negative
value of $d$ subject to the
requirement that $cd$ is a multiple of $N$. This gives
$d=-N$ and hence
\be \label{ebbyd2}
{b\over d} = {1\over N}\, .
\ee

Following this rule we can now construct the circles 
which border
the domain from below. 
For the purpose of illustration we shall carry out the first few steps. 
Using \refb{ebbyd2} we see that the
first circle segment beginning at 0 ends at $1/N$. 
For $N=1$ this completes the story since we have already
reached the
point 1. For $N\ge 2$ we need to proceed further. Taking $p/q=1/N$
we see from \refb{ebbyd} that $b=(d-1)/N$. Since $q=N$, the
condition that $qd$ is a multiple of $N$ is trivially satisfied.
Thus we need to choose $d$ to be the maximum negative number
for which $b=(d-1)/N$ is integer. This gives $d=-(N-1)$ and $b=-1$.
Thus the circle ends at $b/d=1/(N-1)$. For $N=2$ this completes
the story but for $N\ge 3$ we need to proceed further. At the next
stage we begin with $p/q=1/(N-1)$ and get $b=-(N-1)$, 
$d=-N(N-2)$. Thus $b/d=(N-1)/((N(N-2))$. This does not complete
the story for any $N\ge 3$; {\it e.g.}
for $N=3$ this gives $b/d=2/3$. By 
continuing
this process one can show that for $N=3$ we reach the point
1 at the next step via the wall corresponding
to $\pmatrix{a & b\cr c & d}=\pmatrix{2 & -1\cr 3 & -1}$ but
for higher $N$ the story continues further. A schematic diagram
representing these domains for $N=1$, 2 and 3 
have been shown in Fig.~\ref{f1}. As will be discussed in
\S\ref{stest}, for $N\ge 4$ the number of
such walls is infinite. 

This finishes our general analysis of marginal stability walls and
domains bounded by them. Now we focus on one particular domain,
-- the one in which the degeneracy formula given in
\refb{egg1int}, \refb{ep2kk} is valid. 
Since the calculation leading to
\refb{egg1int}, \refb{ep2kk} was performed in a specific corner of the
moduli space, -- the weakly coupled type IIB string theory,
-- all we need to know is how this region is situated 
with respect to the
various marginal stability
walls described here.
However to address this issue we need to 
determine
the relation beween the moduli parameters in the first description,
-- as an orbifold of type IIB string theory, -- and the moduli 
$\tau_\infty$, $M_\infty$ appearing in the BPS mass formula
\refb{esm1}.
We have already stated that
$\tau_\infty=a_\infty+iS_\infty$ 
denotes the asymptotic value of the complex structure
of the torus $(S^1\times \wt S^1)/\ZZZ_N$. We have worked in
a region of the moduli space where it is finite, \i.e.\
$S_\infty$ is neither too large nor too small. The relation between
the other moduli fields in the first description and the matrix valued
moduli field $M$ can be found by following carefully the duality
chain that takes the theory to its second description, 
and using the
identification of $M$ with the geometric quantities in this
description as given in \refb{etenfour}. Let us denote, 
 in the
first description of the theory, 
by $g$ the ten dimensional coupling constant, by 
$(2\pi\sqrt{\alpha'})^4V$ the volume
of $\MM$ measured in the string metric, and by 
$(2\pi\sqrt{\alpha'})^2A$ the area of
$S^1\times \wt S^1/\ZZZ_N$ measured in the string metric. 
Let us also set the anti-symmetric tensor field and
all the Ramond-Ramond fields to zero. Then using the standard
duality transformation rules one can show that the
relevant $4\times 4$ component part of the matrix $M$ that couples
to $Q$ and $P$ are given by
\be \label{ematrix}
M_\infty= \pmatrix{{1/ V} &&& \cr & {g^2/(A^2
 V)} &&\cr
&&  V & \cr &&& {A^2  V/ g^2}}\, .
\ee
Using
the expression for $Q$ and $P$
given in \refb{echvec} one now gets
\ben\label{egets}
Q^T(M_\infty+L)Q &=& {g^2\over A^2  V} \left({n\over N}\right)^2
+ {A^2 V\over g^2} + 2{n\over N}, \nonumber \\
P^T(M_\infty+L)P&=&
{(Q_1-\zeta)^2\over V} +{g^2\over A^2  V} J^2
+  V + 2(Q_1-\zeta), \nonumber \\
P^T(M_\infty+L)Q &=& {g^2\over A^2  V} {nJ\over N} + J\, .
\een
The computation of the degeneracy was 
done in a region where $g$ is small but
$A$ and $V$ are of order 1. Using \refb{egets} 
one can now see that in this region
\be\label{escale3}
Q^T(M_\infty+L)Q >> P^T (M_\infty+L) P, \, 
|P^T (M_\infty+L) Q|\, .
\ee
Even when we deform $M_\infty$ away a little from its diagonal form
\refb{ematrix}, eq.\refb{escale3} continues to hold.

We now study the implication of \refb{escale3} 
on  the location of the 
walls of marginal stability in the $\tau_\infty$ plane.
First consider the case where $cd\ne 0$. In this case we see from
\refb{esm5}, \refb{escale3} that 
\be \label{escale4}
{E\over cd}  \simeq \sqrt{Q^T(M_\infty+L)Q\over
P^T (M_\infty+L) P}>> 1\, .
\ee
Hence for the circle in the $\tau_\infty$ plane described by
\refb{esm4} we have
\be \label{escale5}
|S_\infty| <  {\sqrt{1 + E^2}\over |2cd|} - {E\over 2cd} << 1\, .
\ee
In other words in the $(a_\infty, S_\infty)$ plane the uppermost point
on the circle \refb{esm4} lies little above the $S_\infty=0$ axis and
its center lies deep down in the lower half plane. 
Since $S_\infty\sim 1$ in
the region of the moduli space in which we have worked, we see that
this region lies {\it above} all the circles described by
\refb{esm4} in the $\tau_\infty$ plane.

Next we consider the case $c=0$, -- as discussed earlier the $d=0$
case is equivalent to this. In this case eq.\refb{egets} 
shows that the coefficient of $S_\infty$
in \refb{ecurve1} is small for
finite $b$ and small $g$. Thus for fixed
$M_\infty$ these lines
are almost vertical in the $(a_\infty, S_\infty)$ plane and are
given by
\be \label{evert1}
a_\infty \simeq b\, .
\ee
The $b=0$ line corresponds to the wall of marginal stability
described at the end of \S\ref{sreview}, and as we described in
\S\ref{sreview}, the degeneracy formula actually jumps across
this wall.\footnote{In the analysis of \cite{0605210} the jump 
occured exactly across the line $a_\infty=0$. This can be traced
to the fact that there the analysis was carried out 
using weak coupling approximation
and did not take into account the backreaction
due to switching on $J$. Indeed for $J=0$ we have $P^T(M_\infty+L)
Q=0$, and \refb{ecurve1} for $b=0$ reduces to the vertical line
$a_\infty=0$.} Since our degeneracy formula has been derived in
the small $a_\infty$ region, we see that
the region of validity of our formula is
bounded by the $b=-1$ line on the left, $b=1$ line on the
right and a set of circle segments below.

To summarize, the region of the moduli space  
in which we have carried out our
analysis consists of two domains. One of them, lying between the
$\pmatrix{a & b\cr c & d}=\pmatrix{1 & 0\cr 0 & 1}$ line
and $\pmatrix{a & b\cr c & d}=\pmatrix{1 & 1\cr 0 & 1}$ line
extends to the large $S_\infty$ region, and 
is bounded from below by a set of circles. 
For later reference we 
shall call this the right domain $\RR$ and
denote by $\BB_R$ the set of matrices $\pmatrix{a & b
\cr c & d}$ labelling the boundaries of this domain.
The other domain,
lying between the
$\pmatrix{a & b\cr c & d}=\pmatrix{1 & -1\cr 0 & 1}$ line
and $\pmatrix{a & b\cr c & d}=\pmatrix{1 & 0\cr 0 & 1}$ line
also extends to the large $S_\infty$ region and
is  bounded from below by a set of circles. We shall call this the
left domain $\LL$, and denote by $\BB_L$ the set of matrices
$\pmatrix{a & b
\cr c & d}$ labelling the boundaries of this domain.  We shall
argue in \S\ref{stest} that the
set $\BB_L$ is obtained simply by multiplying the elements
of $\BB_R$ by the matrix $\pmatrix{1 & -1\cr 0 & 1}$ from the
left. 

One important issue that we would like to address is: how does the
degeneracy formula change as we move across a wall of marginal
stability? We already 
know that as we move across the wall \refb{ecurve1} for $b=0$
the degeneracy formula continues to be given by \refb{egg1int}
except for a change in the location of the integration contour
from $\CC$ to $\wh\CC$. We shall argue in
\S\ref{sdualtwo} that this is a general phenomenon; as we move
across any line of marginal stability the 
degeneracy formula will be given
by the same expression except for
a change in the integration contour.  

\sectiono{Duality Transformation of the Degeneracy Formula} 
\label{sdualtwo}

As noted in \S\ref{sreview}, 
the degeneracy formula \refb{egg1int},
\refb{ep2kk} has been 
written in terms
of T-duality invariant combinations $Q^2$, $P^2$ and $Q\cdot P$
although we have derived the formula only for a special class of
charge vectors. In this section we shall discuss what information about
the degeneracy formula can be extracted using the 
T- and S-duality symmetries of the theory.

We begin by studying the consequences of 
the T-duality symmetries of the theory. It follows from
\refb{eag9}, \refb{sadd1} that if a T-duality transformation takes
a charge vector $(\vec Q, \vec P)$ to $(\vec Q',\vec P')$ then
\be \label{sadd2}
Q^{\prime 2}=Q^2, \quad P^{\prime 2}=P^2, \quad Q'\cdot P'
= Q\cdot P\, .
\ee
However there may be
pairs of charge vectors with the same $Q^2$, $P^2$ and
$Q\cdot P$ which are not related by a T-duality transformation.
Clearly T-duality invariance of the theory cannot give us
any relation between the degeneracies associated with
such a pair of charge vectors.
In what follows we shall focus on charge vectors $(\vec Q',\vec P')$
which are in the same T-duality orbit of a charge vector $(\vec Q,
\vec P)$ for which we have derived \refb{egg1int}. 

We have denoted by $\RR$ the right domain of the
region of the moduli space described in \S\ref{smarginal} in which
the original formula for $d(\vec Q, \vec P)$ is valid. It is bounded
by a set of marginal stability walls labelled by 
$\pmatrix{a & b\cr c & d} \in\BB_R$.  Let
$\RR'$ denote the image of
$\RR$ under the T-duality map. 
In this case we
expect $d(\vec Q', \vec P')$ in the region
$\RR'$ to be equal to $d(\vec Q, \vec P)$ given
in \refb{egg1int}:
\ben\label{egg1intrp}
d(\vec Q',\vec P') &=& {1\over N}\, \int _\CC d\wt\rho \, 
d\wt\sigma \,
d\wt v \, e^{-\pi i ( N\wt \rho Q^2
+ \wt \sigma P^2/N +2\wt v Q\cdot P)}\, {1
\over \wt\Phi(\wt \rho,\wt \sigma, \wt v)}\, , \nonumber \\
&=& {1\over N}\, \int _\CC d\wt\rho \, 
d\wt\sigma \,
d\wt v \, e^{-\pi i ( N\wt \rho Q^{\prime 2}
+ \wt \sigma P^{\prime 2}2/N +2\wt v Q'\cdot P')}\, {1
\over \wt\Phi(\wt \rho,\wt \sigma, \wt v)}\, ,
\een
where $\CC$ has been defined in \refb{ep2kk}.
In going from the first
to the second line of \refb{egg1intrp} we have used \refb{sadd2}.

Let us now determine the region $\RR'$.
Since under a T-duality
transformation $M\to \Omega M \Omega^T$, and since $\RR'$
is the image of $\RR$ under this map, $\RR'$
is bounded by walls of marginal stability described in \refb{esm4},
\refb{esm5} with $\pmatrix{a & b\cr c & d}\in \BB_R$
and $M_\infty$ in \refb{esm5}
replaced by $\Omega^{-1} M_\infty (\Omega^T)^{-1}$.
Using \refb{sadd1}
we see that this effectively replaces $(\vec Q, \vec P)$ by $(\vec Q',
\vec P')$ in \refb{esm5}. Thus
$\RR'$ is the region of the upper half plane bounded by the
circles:
\be \label{esm4rp}
\left(a_\infty - {ad+bc\over 2cd}\right)^2 
+ \left( S_\infty +{E'\over 2 cd}
\right)^2 = {1\over 4 c^2 d^2} (1 + E^{\prime 2})\, , 
\quad \pmatrix{a & b\cr c & d}\in \BB_R\, ,
\ee
where
\be \label{esm5rp}
E' \equiv {   cd (Q^{\prime T} (M_\infty + L) Q')
+ ab (P^{\prime T} (M_\infty + L) P') 
-(ad + bc) (P^{\prime
T} (M_\infty + L) Q') \over 
\left[ (Q^{\prime T} (M_\infty + L) Q') (P^{\prime T} (M_\infty + L) 
P')
- (P^{\prime T} (M_\infty + L) Q')^2\right]^{1/2}
}\, .
\ee
\refb{egg1intrp}-\refb{esm5rp} are valid for any charge vector
$(\vec Q',\vec P')$ which can be related to the charge vectors
given in \refb{echvec} via a T-duality transformation.

Next we shall analyze the consequences of S-duality symmetry.
An S-duality transformation changes the vector $(\vec Q, \vec P)$
to another vector $(\vec Q'', \vec P'')$ 
and $\tau$ to $\tau''$ 
via the formul\ae\ \refb{sadd3}, \refb{sadd4}.
Thus if $\RR''$ denotes
the image of the region $\RR$ under the  map \refb{sadd4}, then
S-duality invariance implies that
inside $\RR''$ the degeneracy $d(\vec Q'', \vec P'')$ is given by
the same expression \refb{egg1int} for $d(\vec Q, \vec P)$:
\be\label{sadd5}
d(\vec Q'',\vec P'') = {1\over N}\, \int _\CC d\wt\rho \, 
d\wt\sigma \,
d\wt v \, e^{-\pi i ( N\wt \rho Q^2
+ \wt \sigma P^2/N +2\wt v Q\cdot P)}\, {1
\over \wt\Phi(\wt \rho,\wt \sigma, \wt v)}\, .
\ee
We would like to express the right hand side of \refb{sadd5} in
terms of the vectors $\vec Q''$ and $\vec P''$. For this we define
\be\label{epm}  
\pmatrix{\tilde \alpha &\tilde \beta\cr\tilde \gamma &\tilde \delta} = 
\pmatrix{\delta  & \gamma /N \cr \beta  N & \alpha }\in
\Gamma_1(N)\, .
\ee
and
\be\label{e3.3}
  \pmatrix{\wrh''\cr \ws''\cr \wv''}  \equiv 
  \pmatrix{\wrh''_1+i\wrh''_2\cr \ws''_1+i\ws''_2\cr \wv''_1
  +i\wv''_2}=
  \pmatrix{\tilde \alpha^2& \tilde \beta^2&- 2\tilde 
   \alpha
   \tilde \beta\cr
   \tilde \gamma^2&\tilde \delta^2 & - 
   2\tilde
   \gamma\tilde \delta\cr
    -\tilde \alpha\tilde \gamma& - \tilde \beta\tilde 
   \delta & (\tilde \alpha\tilde \delta + \tilde 
   \beta\tilde \gamma)} \pmatrix{\wrh\cr \ws\cr \wv}\, .
   \ee
Using \refb{sadd3}, \refb{epm},
\refb{e3.3} one can easily verify that
\be\label{e3.4}
e^{-\pi i ( N\wt \rho Q^2
+ \wt \sigma P^2/N +2\wt v Q\cdot P)} = 
e^{-\pi i ( N\wt \rho'' Q^{\prime\prime 2}
+ \wt \sigma'' P^{\prime \prime 2}/N +2\wt v'' 
Q''\cdot P'')}   \, ,
\ee
and
\be\label{e3.5}
d\wt\rho \, 
d\wt\sigma \,
d\wt v = d\wt\rho'' \, 
d\wt\sigma'' \,
d\wt v'' \, .
\ee
Furthermore, with the help of eq.\refb{ewtphitrs}
one can show that\cite{0609109}
\be\label{e3.6}
   \wt\Phi(\wt \rho'',\wt\sigma'',\wt v'') = \wt\Phi(
   \wt \rho,\wt\sigma,\wt v)\, .
\ee
If $\CC''$ denotes the image of $\CC$ under the map \refb{e3.3}
then 
eqs.\refb{e3.4}-\refb{e3.6} allow us to express
\refb{sadd5} as
\be\label{e3.7}
d(\vec Q'',\vec P'') = {1\over N}\, \int _{\CC''} d\wt\rho'' \, 
d\wt\sigma'' \,
d\wt v'' \, e^{-\pi i ( N\wt \rho'' Q^{\prime\prime 2}
+ \wt \sigma'' P^{\prime \prime 2}/N +2\wt v Q''\cdot P'')}\, {1
\over \wt\Phi(\wt \rho'',\wt \sigma'', \wt v'')}  \, .
\ee
To find the location of $\CC''$ we note that under the map
\refb{e3.3} the real parts of $\wrh$, $\ws$ and $\wv$ mix
among themselves and the imaginary parts of 
$\wrh$, $\ws$ and $\wv$ mix among themselves.
The initial contour $\CC$ corresponded to a unit cell of
the cubic lattice in the
($\wrh_1$,$\ws_1$,$\wv_1$) space spanned by the basis vectors
$(1,0,0)$, $(0,N,0)$ and $(0,0,1)$. The unimodular
map \refb{e3.3} transforms this into a different unit cell
of the same lattice. We can now use
the 
shift symmetries
\be\label{eshift}
\wt\Phi(\wt\rho,\wt\sigma,\wt v) =
\wt\Phi(\wt\rho+1,\wt\sigma,\wt v) =
\wt\Phi(\wt\rho,\wt\sigma+N,\wt v) =
\wt\Phi(\wt\rho,\wt\sigma,\wt v+1)\, ,
\ee
which are manifest from \refb{edefwtphi}, to
bring the integration region back to the original unit cell.
Thus $\CC''$ and $\CC$ differ only in the values of the imaginary
parts of $\wrh$, $\ws$ and $\wv$. Using \refb{ep2kk},
\refb{e3.3} we see that for the contour $\CC''$,
\ben \label{sadd7}
    \wt \rho''_2 =\tilde \alpha^2\, M_1 +\tilde \beta^2\, M_2 + 2\tilde 
   \alpha
   \tilde \beta \, M_3 \, ,
   \nonumber \\
    \wt\sigma''_2 =\tilde \gamma^2\, M_1 +\tilde \delta^2 \, M_2 + 
   2\tilde
   \gamma\tilde \delta\, M_3 \, ,
   \nonumber \\
    v''_2 = -\tilde \alpha\tilde \gamma\, M_1 - \tilde \beta\tilde 
   \delta \, M_2 - (\tilde \alpha\tilde \delta + \tilde 
   \beta\tilde \gamma)\, M_3 \, .
\een
Thus $\CC''$ is not identical to $\CC$, -- a fact first noticed in
\cite{appear}.
We could try to deform $\CC''$ back to $\CC$, but in that process
we might pick up contribution from the residues at the poles of
$\wt\Phi(\wt \rho'',\wt \sigma'', \wt v'')$. Thus we see that the
degeneracy formula \refb{e3.7} for $d(\vec Q'', \vec P'')$ is not
obtained by simply replacing $(\vec Q, \vec P)$ by
$(\vec Q'', \vec P'')$ in the
expression for $d(\vec Q, \vec P)$. The integration contour $\CC$
also gets deformed to a new contour $\CC''$.

Let us now analyze the region $\RR''$
of the asymptotic moduli space in
which \refb{e3.7} is valid. This is obtained by taking the image
of the region $\RR$ under the transformation \refb{sadd4}.
To determine this region we need to first study the images of the
curves described in \refb{esm4} in the
$a_\infty-S_\infty$ plane. A straightforward analysis shows that
the image of \refb{esm4} is described by the curve
\be \label{esm4rprp}
\left(a_\infty - {a''d''+b''c''\over 2c''d''}\right)^2 
+ \left( S_\infty +{E''\over 2 c''d''}
\right)^2 = {1\over 4 c^{\prime\prime 2} d^{\prime\prime 2}} (1 + 
E^{\prime\prime 2})\, ,
\ee
where
\be \label{sadd8}
\pmatrix{a'' & b'' \cr c'' & d''} = \pmatrix{\alpha & \beta\cr \gamma
& \delta} \pmatrix{a & b\cr c & d}\, ,
\ee
and
\be \label{esm5rprp}
E'' \equiv { c''d'' (Q^{\prime\prime 
T} (M_\infty + L) Q'')
+ a''b'' (P^{\prime\prime  T} (M_\infty + L) P'') 
 -(a''d'' + b''c'') (P^{\prime\prime
T} (M_\infty + L) Q'') \over 
\left[ (Q^{\prime \prime T} (M_\infty + L) Q'') 
(P^{\prime \prime T} (M_\infty + L) 
P'')
- (P^{\prime \prime T} (M_\infty + L) Q'')^2\right]^{1/2}
}\, .
\ee
It can be easily seen that $\pmatrix{a'' & b'' \cr c'' & d''}$ satisfy
the relations \refb{enorm}, \refb{egen}, and the equivalence
relations \refb{escale1}, \refb{esc12} translate to identical
equivalence relations on $\pmatrix{a'' & b'' \cr c'' & d''}$.
Thus the collection of all the
matrices $\pmatrix{a'' & b'' \cr c'' & d''}$ describes
same set $\AAA$ as the collection of all the
matrices $\pmatrix{a & b\cr c & d}$.

Now recall that the original region $\RR$ was bounded by a
set of marginal stability walls $\left\{\pmatrix{a & b\cr
c & d}\right\}\in\BB_R$. 
Thus the region $\RR''$ 
is bounded by the collection of walls described by  
\refb{esm4rprp}, \refb{esm5rprp} with
$\pmatrix{a'' & b''\cr c'' & d''}\in \pmatrix{\alpha & \beta\cr
\gamma & \delta}\BB_R$.

 At this stage it will be instructive to 
 compare the expressions for the degeneracies 
$d(\vec Q,\vec P)$ and $d(\vec Q'', \vec P'')$. There are two
key differences. First of all although both are represented
by similar looking integrals, the contours of integration are
different in the two cases. Second the region of validity $\RR''$
of
the expression for $d(\vec Q, \vec P)$ is not obtained by simple
replacement of $(\vec Q, \vec P)\to (\vec Q'',\vec P'')$ in the
expression for $d(\vec Q, \vec P)$; 
the original region of validity of our formula
was bounded by a set of marginal stability
walls corresponding to a set of matrices
$\pmatrix{a & b\cr c & d}\in\BB_R $, whereas the new region
of validity of the formula is bounded by 
a set of walls corresponding
to the matrices
$ \pmatrix{a'' & b''\cr c'' & d''}\in \pmatrix{\alpha & \beta\cr
\gamma & \delta}\BB_R$.

Typically the new charge vectors $(\vec Q'', \vec P'')$ are not of
the type \refb{echvec}, but in some cases 
it is possible to obtain the
charge vectors $(\vec Q'', \vec P'')$ by T-duality transformation of
another charge vector $(\vec {\wt Q}, \vec {\wt P})$ 
of the form
given in \refb{echvec} satisfying
 \be\label{eqdefrp}
\wt Q^2 = Q^{\prime\prime 2}, \qquad \wt 
P^2 = P^{\prime\prime 2}, \qquad \wt Q\cdot\wt P
= Q''\cdot P''\, .
\ee
 Then for $d(\vec {\wt Q}, \vec {\wt P})$ our
original formula for degeneracy holds.
Using the results given in \refb{egg1intrp}-\refb{esm5rp}
we can now conclude that in the region $\wt\RR''$ 
bounded by the walls \refb{esm5rprp}  with
\be \label{eww1}
\pmatrix{a'' & b''\cr c'' & d''} \in \BB_R
\, ,
\ee
the
degeneracy $d(\vec Q'', \vec P'')$ is given by 
 \refb{e3.7}
with the contour $\CC''$ replaced by $\CC$. 

Thus we now have expressions for $d(\vec Q'', \vec P'')$ in two
different domains, -- $\wt \RR''$ and $\RR''$. In both regions the
degeneracy is given by an integral. The integrand in both cases are
same, but in one case the integration contour is $\CC$ while in
the other case it is $\CC''$.
This shows that
as we cross the walls of marginal stability to move from the
region $\wt\RR''$ to $\RR''$ the expression
for $d(\vec Q'', \vec P'')$ changes by a modification in the location
of the contour of integration.  

Even though the result
was derived under the assumption that $(\vec Q'', \vec P'')$
can be related to a charge vector of the type \refb{echvec} by
T-duality, it is natural to assume that this phenomenon is more
general. In particular, a natural postulate will be that
{\it as we move from the domain 
$\RR$ corresponding
to the set $\BB_R$ to
another domain corresponding to the set $\pmatrix{\alpha & \beta
\cr \gamma & \delta}\BB_R$, with $\pmatrix{\alpha &\beta\cr
\gamma & \delta}\in\Gamma_1(N)$, 
the expression for $d(\vec Q, \vec P)$
gets modified by a replacement of the integration contour
$\CC$ to the contour $\CC''$ corresponding to
\ben \label{sadd7rp}
    \wt \rho_2 =\tilde \alpha^2\, M_1 +\tilde \beta^2\, M_2 + 2\tilde 
   \alpha
   \tilde \beta \, M_3 \, ,
   \nonumber \\
    \wt\sigma_2 =\tilde \gamma^2\, M_1 +\tilde \delta^2 \, M_2 + 
   2\tilde
   \gamma\tilde \delta\, M_3 \, ,
   \nonumber \\
    v_2 = -\tilde \alpha\tilde \gamma\, M_1 - \tilde \beta\tilde 
   \delta \, M_2 - (\tilde \alpha\tilde \delta + \tilde 
   \beta\tilde \gamma)\, M_3 \nonumber \\
    0\le  \wt\rho_1\le 1, \quad
0\le  \wt\sigma_1\le N, \quad 0\le  \wt v_1\le 1\, ,
\een
$M_1$, $M_2$ and $M_3$ being large but fixed positive
numbers with $M_3<< M_1, M_2$ and
$ \wt\alpha=\delta$, $\wt\beta = \gamma/N$, $\wt\gamma
   =N\beta$, $\wt\delta = \alpha$.}

Can every domain be related to $\RR$ or $\LL$ via S-duality
transformation? 
If this is so then we can
use duality invariance together with the information about
the  contours $\CC$ and $\wh\CC$ appropriate for
$\RR$ and $\LL$, -- as given in \refb{ep2kk} 
and \refb{ecchat},
--  to find out the location of the integration contour in
every other domain. For $N=1,2,3$ we can answer this question
in the affirmative as  follows.  In these cases the matrix
$\pmatrix{a & b\cr c & d}$ labelling any wall can be
represented by an element of $\Gamma_1(N)$ and hence
can be 
related to the matrix $\pmatrix{1 & 0\cr 0 & 1}$ via a
duality transformation of the form \refb{sadd8}.  Since
$\pmatrix{1 & 0\cr 0 & 1}$ describes the boundary between
the domains $\RR$ and $\LL$, 
the two domains bordering the wall $\pmatrix{a & b\cr c & d}$
must be related to the domains $\RR$ and $\LL$ 
by the same duality transformation. Since this can be done for
every wall of marginal stability, we see that every domain
must be related to either $\RR$ or $\LL$ via a duality
transformation.

For $N\ge 4$ not all the walls can be related to the wall
$\pmatrix{1 & 0\cr 0 & 1}$ via duality transformation.
Nevertheless it may  still be possible to
relate all the domains to $\RR$ and $\LL$ via duality 
transfomation, with the walls not related to 
$\pmatrix{1 & 0\cr 0 & 1}$ being related to some other wall
of $\RR$ or $\LL$. We shall not attempt to settle this
issue here.

Given two contours on two sides of a marginal
stability wall,
if we try to deform one to the other then typically
we shall encounter poles of the integrand, and the two results
will differ by the sum of the residues at these poles. To get a physical
insight into how much this change is as we move across a
given wall, let us consider the wall
corresponding to the matrix $\pmatrix{1 & 0\cr 0 & 1}$. 
This is the wall separating the regions 
$\RR$ and $\LL$, and 
across this wall the original dyon 
becomes unstable against
decay into states with charges 
$(\vec Q, 0)$ and $(0,\vec P)$.
Since across this wall the contour changes from constant
negative $\wt v_2$ to constant
positive $\wt v_2$, we pick up the residue at the pole
$\wv=0$. Now near $\wt v=0$ the function $\wt\Phi$ has the
behaviour\cite{0510147,0602254,0607155,0609109}
\be \label{enearpole}
\wt\Phi(\wrh,\ws,\wv) \simeq -4\pi^2 \wt v^2 \, f(N\wt\rho) 
g(\wt\sigma/N) + \OO(\wv^4)
\, ,
\ee
where $f(\tau)$ and $g(\tau)$ are two functions which have the
interpretation of inverse of partition functions associated with
electric and magnetic half BPS states of the 
theory.
Performing the integral over $\wt v$ in \refb{egg1int}
around the $\wt v=0$ point now gives
\be \label{echange}
-(Q\cdot P) \left\{\int_0^1 d\wrh e^{-i\pi  N \wrh Q^2} 
\left( f(N\wrh)\right)^{-1}\right\}  \left\{{1\over N} \int_0^N
d\ws e^{-i \pi  \ws P^2/N} 
\left( g(\ws/N)\right)^{-1}\right\}\, .
\ee
This formula can be given a simple physical interpretation.
The second and the third factors represent the
degeneracies of the electric and magnetic half BPS states
into which the original dyon decays on the marginal stability
wall. The $Q\cdot P$ factor on the other hand is associated
with the supersymmetric quantum mechanics describing the
relative motion of the electric and the magnetic 
system\cite{0605210}.
In particular it represents the number of states whose binding
energy vanishes as we reach the marginal
stability wall\cite{pope,9912082}. Thus this is the number
of states which disappear from the spectrum as we cross the
wall.

\sectiono{The Large Charge Limit} \label{sblack}

In this section we shall argue that even though the complete
spectrum changes discontinuously when the asymptotic value
of the axion field changes sign, the large charge expansion
is not affected by this change. 
Our starting point is the result derived in
\cite{0605210,0607155,0609109} 
that the poles in the integrand in the expression for 
$d(\vec Q,\vec P)$ come
from the  second order zeroes of $\wt\Phi$  at\footnote{For $\MM=K3$
this result can be found in appendix E of \cite{0605210}. For
the general case the set of all the zeroes and poles of $\wt\Phi$ were
listed in \cite{0609109}, but we did not attempt to separate out
the zeroes from the poles. A careful analysis shows that the only
zeroes come from the set \refb{eonly}.}
\ben\label{eonly}
&&   n_2 ( \ws \wrh  -\wv ^2) + j\wv  + 
n_1 \ws  -m_1 \wrh  + m_2
 =0\, , \nonumber \\
 && 
m_1 n_1 + m_2 n_2 +\frac{j^2}{4} = {1\over 4} \, ,\nonumber \\
&& m_1\in N\ZZZ, 
\quad n_1, m_2, n_2 \in \ZZZ, \quad j\in 2\ZZZ+1
\, . 
\een
For large charges
the leading contribution to the degeneracy comes from
the pole 
at $n_2=1$\cite{9607026,0412287,0510147,0605210,
0607155,0609109}. Contribution from the poles with
$n_2\ne 1$ are exponentially suppressed compared to the
leading contribution.

Consider now the contours $\CC$ and $\CC''$ given in
\refb{ep2kk} and \refb{sadd7rp} respectively. 
Both contours have the same range of integration
over the real parts of $\wrh$, $\ws$ and $\wv$
\be \label{erho1}
0\le\wrh_1<1, \qquad 0\le\ws_1< N, \qquad
0\le \wv<1\, .
\ee
However they
differ in the values of $\wrh_2$, $\ws_2$ and $\wv_2$.
We have for
\ben\label{etwocontour}
\CC &:&  \wrh_2=M_1, \quad \ws_2=M_2, \quad \wv_2=-M_3\, ,
\nonumber \\
\CC'' &:&  \wrh_2=\tilde\alpha^2\, M_1 
+\tilde \beta^2\, M_2 + 2\tilde 
   \alpha
   \tilde \beta \, M_3, \quad \ws_2=\tilde \gamma^2\, M_1 
   +\tilde \delta^2 \, M_2 + 
   2\tilde
   \gamma\tilde \delta\, M_3, \nonumber \\
   && \wv_2=
   -\tilde \alpha\tilde \gamma\, M_1 - \tilde \beta\tilde 
   \delta \, M_2 - (\tilde \alpha\tilde \delta + \tilde 
   \beta\tilde \gamma)\, M_3 \, ,
\een
where $M_1$, $M_2$ and $M_3$ are fixed 
numbers with $M_1,M_2>>1$ and
$|M_3|<<M_1, M_2$.
Our goal will be to show that we can deform the contour $\CC$
to $\CC''$ without hitting any of the poles given in
\refb{eonly} except those with $n_2=0$. Since for large charges
the contribution
to the degeneracy comes from the poles at $n_2=1$
up to
exponentially suppressed terms, this will show that
the change in the degeneracy due to the change in the contour
of integration is exponentially suppressed compared to the
leading contribution.

To proceed we shall choose a specific path along
which we deform the contour. We take this to be along the
straight line
joining the points $(M_1, M_2, M_3)$ and 
$(\tilde\alpha^2\, M_1 
+\tilde \beta^2\, M_2 + 2\tilde 
   \alpha
   \tilde \beta \, M_3, \tilde \gamma^2\, M_1 
   +\tilde \delta^2 \, M_2 + 
   2\tilde
   \gamma\tilde \delta\, M_3,
-\tilde \alpha\tilde \gamma\, M_1 - \tilde \beta\tilde 
   \delta \, M_2 - (\tilde \alpha\tilde \delta + \tilde 
   \beta\tilde \gamma)\, M_3)$ in the $(\wrh_2, \ws_2,\wv_2)$
   space. Points on this line can be parametrized by a real
   number $\lambda$ lying between 0 and 1, with
\ben \label{estrai}
\wrh_2 &=& M_1 + \lambda\, \{ (\tilde\alpha^2-1)\, M_1 
+\tilde \beta^2\, M_2 + 2\tilde 
   \alpha
   \tilde \beta \, M_3\} \nonumber \\
\ws_2 &=& M_2 + \lambda \, \{ \tilde \gamma^2\, M_1 
   +(\tilde \delta^2-1) \, M_2 +
   2\tilde
   \gamma\tilde \delta\, M_3\} \nonumber \\
\wv_2 &=& -M_3 +\lambda \, \{ -\tilde \alpha\tilde \gamma\, M_1 
- \tilde \beta\tilde 
   \delta \, M_2 - (\tilde \alpha\tilde \delta + \tilde 
   \beta\tilde \gamma-1)\, M_3\}\, .
\een
For each $\lambda$ we take the real parts of $\wrh$, $\ws$ and $\wv$ 
to be
in the range \refb{erho1}. 

We want to show that \refb{eonly} and \refb{estrai} have no
simultaneous solutions except for $n_2=0$. For this we write down
separately the real and imaginary parts of eq.\refb{eonly}, as
well as the constraint on $\vec m$, $\vec n$, $j$:
\be \label{esep1}
\ws_2 \wrh_2 - \wv_2^2 - \ws_1\wrh_1 + \wv_1^2
-{j\over n_2} \wv_1 - {n_1\over n_2} \ws_1
+ {m_1\over n_2} \wrh_1 -{m_2\over n_2} = 0\, ,
\ee
\be \label{esep2} 
n_2 (\ws_1 \wrh_2 + \ws_2 \wrh_1 -2 \wv_1 \wv_2)
+ j\wv_2 +n_1\ws_2-m_1 \wrh_2 = 0\, ,
\ee
\be \label{esep3}
m_1 n_1 + m_2 n_2 +{j^2\over 4}={1\over 4}\, .
\ee
We now eliminate $m_1$ and $m_2$ using \refb{esep2}, 
\refb{esep3}
to write  \refb{esep1} as
\be \label{esep4}
 (\ws_2 \wrh_2 - \wv_2^2) + \left[\pmatrix{j & n_1}\, A
\, \pmatrix{j\cr n_1} + aj+ bn_1 + c\right]=0\, ,
\ee
where
\be \label{esep5}
A = {1\over 4 n_2^2} 
 \pmatrix{1 &  {2\wv_2/
\wrh_2} \cr 
2\wv_2/
\wrh_2 & {4\ws_2/\wrh_2}}\, ,
\ee
\be \label{esep6}
a = {1\over  n_2}\, \left({\wv_2 \over \wrh_2}\wrh_1 
-\wt v_1\right), \quad
b = 2{\ws_2\over n_2\wrh_2} \left(\wrh_1 - {\wv_2\over \ws_2}\wv_1
\right), \quad c = v_1^2   
+ \wrh_1 \left( {\ws_2\over \wrh_2} \wrh_1
- 2{\wv_2\over \wrh_2} \wv_1 \right)-{1\over 4 n_2^2}\,.
\ee
Now for our contour $\wrh_1$, $\ws_1$ and $\wv_1$ are always finite.
For $M_1$ and $M_2$  large and of the same order with 
$|M_3|<<M_1,M_2$,  the quantities
$\wrh_2$ and $\ws_2$  given in \refb{estrai} always remain large
and positive, and the ratios $\ws_2/\wrh_2$, $\wv_2/\wrh_2$,
$\wv_2/\ws_2$ etc. are bounded from above by finite numbers.
Thus the quantities $a$, $b$ and $c$ remain finite.
On the other hand in this limit we have 
\be \label{esep7}
\wrh_2\ws_2 - \wv_2^2 \simeq \lambda (1-\lambda) 
\left( \tilde\gamma^2 M_1^2 + \tilde\beta^2 M_2^2 \right)
+ \left( \lambda^2 + (1-\lambda)^2 +  \lambda (1-\lambda)
(\tilde\alpha^2 +\tilde\delta^2)\right) M_1 M_2\, .
\ee
In the range $0\le\lambda\le 1$ each term in this expression is
non-negative, and  \refb{esep7}
remains large and positive, -- of order $M_1M_2$, -- 
in the limit
of large $M_1$, $M_2$. This also shows that the matrix
$A$ defined in \refb{esep5} is nondegenerate in this limit and in
fact has finite positive eigenvalues. As a result the term in the square
bracket in \refb{esep4} is bounded from below by a finite number
\be \label{efinite}
c- {1\over 4} \pmatrix{a & b} \, A^{-1}\, \pmatrix{a\cr b}\, ,
\ee 
and can never cancel the first term in \refb{esep4}
for any value of $j$ and $n_1$
for sufficiently large $M_1$ and $M_2$.
Hence \refb{esep4} cannot be satisfied.
This shows that it is impossible to find a similtaneous solution to the
eqs.\refb{eonly} and \refb{estrai} for $n_2\ne 0$, 
and hence the contour $\CC$ can
be deformed to $\CC''$ along the path \refb{estrai} without
encountering the poles of the integrand given in \refb{eonly}
for $n_2\ne 0$.

This establishes that in the limit of large charges the degeneracy
remains  the same in different domains in the moduli space up
to nonperturbative terms.
This result is consistent with the fact that for a black hole of
charge $(\vec Q, \vec P)$ in this theory  we have a
stable supersymmetric attractor for $P^2>0$, $Q^2>0$,
$P^2 Q^2 > (Q\cdot P)^2$. Thus the near horizon geometry of these
black holes is always given by this attractor point and is independent
of the asymptotic moduli even if this requires the attractor flow
to cross one or more walls of marginal stability.

\sectiono{Test of S-duality Invariance} \label{stest}

In \S\ref{sdualtwo} we used S-duality invariance to determine the
locations of the integration contour in the degeneracy formula in
different domains in the moduli space. However this did not provide
a test of S-duality. In this section we shall describe some tests of
S-duality that one could perform.
\begin{enumerate}
\item If there is an S-duality transformation that leaves the set
$\BB_R$ invariant, then under such a transformation the contour
$\CC$ either should not transform, or should transform to another
contour that is continuously deformable to $\CC$ without passing
through any poles.
\item Analysis of \cite{0605210} has shown that inside the 
left domain
$\LL$ corresponding to the set of matrices $\BB_L$, the degeneracy
is given by performing integration over the
contour $\wh\CC$ described in \refb{ecchat}.
Thus if there is an S-duality transformation that maps the set
$\BB_R$ to the set $\BB_L$ then such a transformation must
map the contour $\CC$ to the contour $\wh\CC$ or another
contour deformable to $\wh\CC$ without passing through
any pole.
\end{enumerate}
In fact for all values of $N$ 
there is an S-duality transformation that maps
$\BB_R$ to $\BB_L$. It is given by
\be \label{egiven}
\pmatrix{\alpha & \beta \cr \gamma & \delta}
= \pmatrix{1 & -1\cr 0 & 1}\, .
\ee
To see this we note that it maps the $\pmatrix{1 & 0\cr 0 & 1}
\in\BB_R$ to $\pmatrix{1 & -1\cr 0 & 1}\in \BB_L$ and
$\pmatrix{1 & 1\cr 0 & 1}
\in\BB_R$ to $\pmatrix{1 & 0\cr 0 & 1}\in \BB_L$. Since
two domains
sharing two common boundaries must be identical, the action
of \refb{egiven} must map $\BB_R$ to $\BB_L$.
Using this transformation we can convert all tests of the first type into
tests of the second type; all we need to do is to left multiply the duality
transformation preserving $\BB_R$ by \refb{egiven} to construct
a duality transformation that maps $\BB_R$ to $\BB_L$.

We shall now try to verify that 
the transformation \refb{egiven} maps $\CC$ to $\wh\CC$ or a 
contour deformable to $\wh\CC$ without passing through any poles.
Using \refb{epm}, \refb{sadd7rp} we see that this 
transformation maps the
contour $\CC$ given in \refb{ep2kk} to
\ben \label{emaps}
\wrh_2 = M_1, \qquad \ws_2 = N^2 M_1 + M_2 - 2 N M_3, \qquad
\wv_2 = NM_1 -M_3\, , \nonumber \\
0\le  \wt\rho_1\le 1, \quad
0\le  \wt\sigma_1\le N, \quad 0\le  \wt v_1\le 1\, .
\een
This is different from $\wh\CC$ given in
\refb{ecchat} for which $(\wrh_2,\ws_2,\wv_2)
=(M_1,M_2,M_3)$. Thus we need to verify that
we can deform the contour \refb{emaps} to
$\wh\CC$  without encountering
any pole. {}From the analysis in \S\ref{sblack} we already know that
the poles at \refb{eonly} for $n_2\ne 0$ are not encountered; thus
we need to look for poles with $n_2=0$. They occur at
\be\label{eonlysp}
\left( j\wv  + 
n_1 \ws  -m_1 \wrh  + m_2
\right)=0\, , 
\ee
with 
\be \label{enra}
m_1\in N\ZZZ,
\quad n_1, m_2  \in \ZZZ, \quad j\in 2\ZZZ+1, \quad
m_1 n_1   +\frac{j^2}{4} = {1\over 4}\, . \nonumber \\
\ee
Taking the imaginary part of eq.\refb{eonlysp} we get
\be \label{eks1}
 j\wv_2  + 
n_1 \ws_2  -m_1 \wrh_2 = 0\, .
\ee
For fixed $j$, $n_1$, $m_1$
this describes a plane in the $(\wrh_2,\ws_2,\wv_2)$ space.
Our job is to show that points in the $(\wrh_2,\ws_2,\wv_2)$ space
given in \refb{ecchat} and \refb{emaps} lie on the
same side of this plane so that we can deform them to each other
without going through this plane. For this we need to show that
the left-hand side of \refb{eks1}, evaluated at \refb{ecchat} and
\refb{emaps} have the same sign:
\be \label{eks2}
(jM_3 + n_1 M_2 -m_1 M_1) \left(
j (NM_1 -M_3) + n_1 (N^2 M_1 + M_2 - 2 N M_3)
- m_1 M_1
\right) >0\, .
\ee
We can simplify the analysis by setting $M_1=M_2=M$. In this
case the left hand side of \refb{eks2} has the form
\ben \label{eks3}
&&  (n_1 - m_1) \{Nj + (N^2 + 1) n_1 - m_1\} M^2 
\nonumber \\
&& + \{(Nj + (N^2 + 1) n_1 - m_1) j
 -(n_1 - m_1) ( j + 2 N n_1) \} \, M \, M_3 \nonumber \\
&& - j (j+2Nn_1) M_3^2 \, .
\een
Using \refb{enra} the coefficient of the $M^2$ term can be
brought to the form
\be \label{em2}
\left( {Nj\over 2} + n_1 - m_1\right)^2 + N^2 n_1^2 - 
{N^2\over 4}\, .
\ee
If $n_1\ne 0$ then this is strictly positive. If $n_1=0$ then
from \refb{enra} we have $j=\pm 1$. We shall choose $j=1$ by
using the freedom of changing the signs of $n_1$, $m_1$,
$n_2$, $m_2$ and $j$
simultaneously without changing the location of the pole.
Since $m_1\in N\ZZZ$,
in this case \refb{em2} is
strictly positive if $m_1\ne 0,N$, and vanishes for $m_1=0,N$.

Now as long as \refb{em2} is non-vanishing and positive,
we can make the first term of \refb{eks3} dominate over others
by taking $M$ to be arbitrarily large, and hence \refb{eks3}
is positive as required. Thus we only need to worry
about is the case $n_1=0$, $j=1$, $m_1=0,N$ when the order
$M^2$ term vanishes. In both cases \refb{eks3} takes the
form:
\be \label{em3}
N\, M\, M_3 - M_3^2\, .
\ee
Since $M$ is large and positive, and $M>>M_3>0$, this is strictly
positive. This shows that \refb{eks2} holds for all
$j$, $m_1$, $n_1$ and hence we do not encounter any pole while
deforming the contour \refb{emaps} to \refb{ecchat}.

For the special case of $N=1$ one can consider another map that
takes us from the set $\BB_R$ to $\BB_L$. It is
\be \label{em4}
\pmatrix{\alpha & \beta \cr \gamma & \delta} =
\pmatrix{0 & 1\cr -1 & 0}\, .
\ee
Under this map the vertices 0, 1 and $i\infty$ of the domain $\RR$
get mapped to the points $i\infty$, $-1$ and $0$ respectively.
The latter set is precisely the vertices of the domain $\LL$.

Using \refb{epm}, \refb{sadd7rp} 
we see that this transformation maps the
original contour to\cite{appear}
\ben \label{em5}
\wt \rho_2=M_2, \quad  \wt\sigma_2 = M_1, \quad
 \wt v_2 = M_3, \nonumber \\
 0\le  \wt\rho_1\le 1, \quad
0\le  \wt\sigma_1\le N, \quad 0\le  \wt v_1\le 1\, .
 \een
 This exactly coincides with \refb{ecchat} up to an exchange
 of $M_1$ and $M_2$. Since exchange of $M_1$ and $M_2$
 does not change the integral (we could take $M_1=M_2$)
 we see that
 this constraint of S-duality is satisfied trivially.
 
 Returning to the case of general $N$, one can identify the
 following additional $\Gamma_1(N)$
 transformation 
 that maps the set $\BB_R$ to $\BB_L$:
 \be \label{ede1}
 \pmatrix{\alpha & \beta \cr \gamma & \delta}
 = \pmatrix{1 & 0\cr -N & 1}\, .
 \ee
 This maps the contour $\CC$ to
 \ben \label{ede2}
\wrh_2 = M_1+M_2-2M_3, 
\qquad \ws_2 = M_2, \qquad
\wv_2 = M_2 -M_3\, , \nonumber \\
0\le  \wt\rho_1\le 1, \quad
0\le  \wt\sigma_1\le N, \quad 0\le  \wt v_1\le 1\, .
\een
Consistency with duality invariance requires that we must be
able to deform this contour to $\wh\CC$ without encountering
any pole.
One can proceed to analyze this exactly in the same manner as
we did for \refb{emaps} and arrive at the condition that 
in order to be able to deform this contour to $\wh\CC$ 
given in \refb{ecchat} we need
the following quantity to be positive (analog of eq.\refb{eks3}):
\ben \label{ede3}
&& (n_1-m_1) (n_1 - 2 m_1 +j)M^2 \nonumber \\
&& + \{ j(n_1 - 2m_1+j) + (n_1-m_1) (2 m_1-j)\} MM_3
\nonumber \\
&& + j (2m_1 -j) M_3^2\, .
\een
Using \refb{enra} the coefficient of the $M^2$ term can be
brought to the form
\be \label{ede4}
\left({j\over 2} + n_1 - m_1\right)^2 + m_1^2 -{1\over 4}\, .
\ee
If $m_1\ne 0$ it is strictly positive. For $m_1=0$ from
\refb{enra} we have $j=1$. 
 Since $n_1\in \ZZZ$,
in this case \refb{ede4} is
strictly positive if $n_1\ne 0,-1$, and vanishes for $n_1=0,-1$.
As long as \refb{ede4} is non-vanishing and positive,
we can make the first term of \refb{ede3} dominate over others
by taking $M$ to be arbitrarily large, and hence \refb{ede3}
is positive as required. Thus we only need to worry
about is the case $m_1=0$, $j=1$, $n_1=0,-1$ when the order
$M^2$ term vanishes. In both cases \refb{ede3} takes the
form:
\be \label{ede5}
M\, M_3 - M_3^2\, .
\ee
Since $M$ is large and positive, and $M>>M_3>0$, this is strictly
positive. This shows that \refb{ede3} is positive for all
$j$, $m_1$, $n_1$ and hence we do not encounter any pole while
deforming the contour \refb{ede2} to \refb{ecchat}.

If we consider the $\Gamma_1(N)$ element 
\be \label{ede6}
g_0= \pmatrix{1 & 1\cr 0 & 1} \pmatrix{1 & 0 \cr -N & 1}
= \pmatrix{1-N & 1 \cr -N & 1}\, ,
\ee
then it clearly leaves the set $\BB_R$ unchanged since it is given
by a map from $\BB_R$ to $\BB_L$ followed by the inverse of
a map from $\BB_L$ to $\BB_R$.
Our results establish that under this transformation 
the contour $\CC$ is
mapped to another contour that is continuously deformable to $\CC$.
This is turn establishes that any power of $g_0$ will also have the same
property. We can now follow this up with the transformation
\refb{egiven} to construct a set of duality transformations that maps
the set $\BB_R$ to $\BB_L$ and maps the contour $\CC$ to another
contour deformable to $\wh \CC$, thereby providing a test of the
corresponding duality transformtion. 
For $N=1$ this includes in particular
the element \refb{em4} considered earlier.

Can every element of $\Gamma_1(N)$ that preserves the set $\BB_R$
be expressed as a positive or negative power of $g_0$? If so
then our test of S-duality invariance of the degeneracy formula
would be
complete. To address this issue
note that a $\Gamma_1(N)$ 
transformation that maps $\BB_R$ to $\BB_R$ must take
adjacent walls to adjacent walls. Furthermore the map must
be orientation preserving. Thus the action of any such
transformation on $\BB_R$ must preserve the cyclic ordering
of the walls and vertices. The map $g_0$ 
indeed has this property. It moves
the walls and vertices by two steps clockwise, taking 0 to 1,
$i\infty$ to $1 - {1\over N}$, 1 to $1 - {1\over N-1}$ etc.
$g_0^{-1}$ causes a shift by two steps in the anti-clockwise
direction. Powers of $g_0$ will move the walls clockwise or
anti-clockwise by even number of steps. 
Are there elements of $\Gamma_1(N)$ which
shift the walls by odd number of steps? If so then by
combining it with appropriate positive or negative powers of
$g_0$ we can generate a transformation that
shifts every wall of $\BB_R$ by
one step in the clockwise direction. Such a move will map
0 to $i\infty$, $i\infty$ to 1 and 1 to $1-{1\over N}$. The unique
SL(2,R) map that implements this is the matrix
$\pmatrix{\sqrt N & -1/\sqrt N\cr \sqrt N & 0}$. This is 
clearly not an element of $\Gamma_1(N)$ for any $N$ other
than $N=1$.

This would seem to indicate that all elements of
$\Gamma_1(N)$ preserving $\BB_R$ are generated by
$g_0$. However there is an
additional subtlety arising out of the fact that for $N> 4$
the element $g_0$ has a pair of fixed points on the real line
corresponding to\footnote{For $N=4$ there is a single fixed point
and hence it does not divide the vertices into two sets. Thus every
pair of vertices separated by even number of steps can still be
related by a $g_0$ transformation.}
\be \label{efi1}
\tau_\infty={1\over 2} \left( 1 \pm \sqrt{1 -{4\over N}}\right)\, .
\ee
These provide accumulation points of the vertices of $\RR$; indeed
if we begin with any vertex of $\RR$ 
and apply $g_0$ or $g_0^{-1}$ transformation
successively to generate other vertices they accumulate at the
points given in \refb{efi1}. As a result there are infinite
number of walls bordering $\RR$, and
a wall or a vertex that is situated within the
range
\be \label{efi2}
{1\over 2} \left( 1 - \sqrt{1 -{4\over N}}\right) <\tau_\infty
< {1\over 2} \left( 1 + \sqrt{1 -{4\over N}}\right) \, ,
\ee
can never be related to a
wall or a vertex lying outside this range by a $g_0$ transformation.
This opens up the possibility that there may be additional
elements of $\Gamma_1(N)$ which map the walls and vertices
outside the range \refb{efi2} to walls and vertices
inside this range and vice versa, preserving the cyclic ordering.

To examine this issue we need to first identify 
some 
vertices lying within the range \refb{efi2}. 
For this we shall first prove that the point $1/2$ must be a vertex.
If it is not a vertex then there must be a wall that goes over 1/2; by
symmetry of the problem the vertices at the two ends of this wall must
be
situated symmetrically about 1/2. Let us take them to be
\be \label{epq1}
{1\over 2} \pm {p\over q} = {q\pm 2p\over 2q}\, ,
\ee
with $p,q$ relatively prime. This wall will correspond to a
matrix $\pmatrix{a & b\cr c & d}$ with $a/c=(q-2p)/2q$ and
$b/d = (q+2p)/2q$. Now we see that if $q$ is odd then 
$(q\pm 2p)$ are odd and hence both $c$ and $d$ must be divisible
by 2. This of course is incompatible with the relation $ad-bc=1$.
If on the other hand $q$ is even (say $2m$) then $p$ must be
odd and we can express \refb{epq1} as
\be \label{epq2}
{m\pm p\over 2m}\, .
\ee
Thus we have $a/c=(m-p)/2m$ and $b/d=(m+p)/2m$. $m$ must be
odd so that $(m\pm p)$ are even; otherwise we again run into the
problem of both $c$ and $d$ being even. Now since
$c$ and $d$ cannot have a common factor, between $m+p$ and
$m-p$ they should be able to cancel all the factors in $2m$. 
This in particular will mean that their product
$(m^2-p^2)$ must be divisible by $m$. Thus $p^2$ should be divisible
by $m$. This is in contradiction with the assumption that $p$
and $q$ (and hence $p$ and $m$) are relatively prime except for
$m=1$. The latter corresponds to the wall connecting 0 and 1
and is not relevant for us.

This shows that $1/2$ must be a vertex (or accumulation point
of vertices) of $\RR$. We can now begin with
${1\over 2}$ and identify the vertices to the left and right
of this using the analysis described below \refb{ebbyd}. For odd
$N$ this gives three points
\be\label{efi3}
{1\over 2} - {1\over 2N}, \quad {1\over 2}, \quad {1\over 2}
+{1\over 2N}\, ,
\ee
in ascending order. On the other hand for $N=2M$ with $M$ odd
we get the three points to be
\be\label{efi4}
{1\over 2} - {1\over 2M}, \quad {1\over 2}, \quad {1\over 2}
+{1\over 2M}\,.
\ee
If $N$ is a multiple of 4 then
one can show that there is no wall ending
at 1/2 and hence 1/2 must be an accumulation point. We shall not
deal with this case here.
We can now ask how $g_0$ acts on these vertices. For $N=5$
and $N=6$ one can check that it moves the left-most vertex
in the set \refb{efi3} or \refb{efi4} to the right-most vertex. By
continuity we can conclude that in the range \refb{efi2},
$g_0$ moves every vertex and wall by two steps in the
{\it anti-clockwise} direction. There is no contradiction with
the fact that outside the range \refb{efi2} $g_0$ moves the points
in the clockwise direction since the boundaries of this region are
fixed points of $g_0$. Thus every wall in the range \refb{efi2}
can be related by $g_0$ action
to one of the two walls connecting the vertices
given in \refb{efi3} or \refb{efi4} for $N=5,6$.
For $N\ge 7$ there are more than two
vertices between a given vertex and its $g_0$ image and the situation
is more complicated. 

Restricting ourselves to the cases $N=5,6$ we now ask the following
question: is there a $\Gamma_1(N)$
transformation that maps the vertices lying 
inside the range \refb{efi2} to
vertices outside
the range \refb{efi2}? We focus on the possible action of this map
on the three vertices 
given in \refb{efi3} or
\refb{efi4}. If such a transformation 
exists then by left multiplying it with powers of
$g_0$ we can always bring the three final points to
1, $i\infty$, 0 or $i\infty$, 0, $1/N$. Examining the two cases
separately we find that there is no such $\Gamma_1(N)$ transformation
mapping \refb{efi3} or
\refb{efi4}  to 
these points (although for odd $N$ there is a $\Gamma_0(N)$ 
transformation
$ \pmatrix{2 & -1\cr N & -(N-1)/ 2} $
which takes the three points
given in \refb{efi3} to 
$i\infty$, 0 and $1/N$). 

This finally establishes that 
for $N\le 6$, all the elements
of $\Gamma_1(N)$ which preserve the set $\BB_R$ are obtained
by taking positive or negative powers of $g_0$. Since we have
checked that these transformations take the contour $\CC$ to
another contour deformable to $\CC$ without encountering any
poles, this completes our test of S-duality invariance for $N\le 6$.

 \bigskip

{\bf Acknowledgement:} I would like to thank Atish Dabholkar,
Justin David, Tohru Eguchi, 
Davide Gaiotto, Dileep Jatkar, Suresh Nampuri and
Hirosi Ooguri for
useful discussions. This work was supported  generously
by the people of India, J.C. Bose fellowship of the
Department of Science and Technology of Govt. 
of India and Moore distinguished scholarship at
California Institute of Technology.


\begin{thebibliography}{99}

 

\bibitem{0605210}
  J.~R.~David and A.~Sen,
  ``CHL dyons and statistical entropy function from D1-D5 system,''
  arXiv:hep-th/0605210.

\bibitem{0607155}
  J.~R.~David, D.~P.~Jatkar and A.~Sen,
  ``Dyon spectrum in N = 4 supersymmetric type II string theories,''
  arXiv:hep-th/0607155.

\bibitem{0609109}
  J.~R.~David, D.~P.~Jatkar and A.~Sen,
  ``Dyon spectrum in generic N = 4 supersymmetric Z(N) orbifolds,''
  arXiv:hep-th/0609109.
  
\bibitem{9607026}
R.~Dijkgraaf, E.~P.~Verlinde and H.~L.~Verlinde,
``Counting dyons in N = 4 string theory,''
Nucl.\ Phys.\ B {\bf 484}, 543 (1997)
[arXiv:hep-th/9607026].

\bibitem{0510147}
  D.~P.~Jatkar and A.~Sen,
  ``Dyon spectrum in CHL models,''
  JHEP {\bf 0604}, 018 (2006)
  [arXiv:hep-th/0510147].

\bibitem{0602254}
  J.~R.~David, D.~P.~Jatkar and A.~Sen,
  ``Product representation of dyon partition function in CHL models,''
  JHEP {\bf 0606}, 064 (2006)
  [arXiv:hep-th/0602254].

\bibitem{0505094}
D.~Shih, A.~Strominger and X.~Yin,
``Recounting dyons in N = 4 string theory,''
arXiv:hep-th/0505094.

\bibitem{0506249}
D.~Gaiotto,
``Re-recounting dyons in N = 4 string theory,''
arXiv:hep-th/0506249.

\bibitem{0612011}
  A.~Dabholkar and D.~Gaiotto,
  ``Spectrum of CHL dyons from genus-two partition function,''
  arXiv:hep-th/0612011.
  
\bibitem{0603066}
  A.~Dabholkar and S.~Nampuri,  
  ``Spectrum of dyons and black holes in 
  CHL orbifolds using Borcherds lift,''
  arXiv:hep-th/0603066.

\bibitem{9712211}
  O.~Bergman,
   ``Three-pronged strings and 1/4 BPS states in N=4 super-Yang-Mills
  theory,''
  Nucl.\ Phys.\ B {\bf 525}, 104 (1998)
  [arXiv:hep-th/9712211].
  
\bibitem{9804160}
  O.~Bergman and B.~Kol,
  ``String webs and 1/4 BPS monopoles,''
  Nucl.\ Phys.\  B {\bf 536}, 149 (1998)
  [arXiv:hep-th/9804160].

\bibitem{appear}
A.~Dabholkar, D.~Gaiotto and S.~Nampuri, private communications
and to appear.

\bibitem{0010222}
F.~Denef,
``On the correspondence between D-branes 
and stationary supergravity solutions of type
II Calabi-Yau compactifications'', 
arXiv:hep-th/0010222.

\bibitem{0101135}
F.~Denef,
``Split attractor flows and the spectrum of 
BPS D-branes on the Quintic'', 
arXiv:hep-th/0101135.

\bibitem{9207016}
  J.~Maharana and J.~H.~Schwarz,
  ``Noncompact symmetries in string theory,''
  Nucl.\ Phys.\ B {\bf 390}, 3 (1993)
  [arXiv:hep-th/9207016].

\bibitem{9402002}
  A.~Sen,
  ``Strong - weak coupling duality in four-dimensional string theory,''
  Int.\ J.\ Mod.\ Phys.\ A {\bf 9}, 3707 (1994)
  [arXiv:hep-th/9402002].

\bibitem{0508042}
  A.~Sen,
  ``Entropy function for heterotic black holes,''
  JHEP {\bf 0603}, 008 (2006)
  [arXiv:hep-th/0508042].

\bibitem{9511222}
  M.~Bershadsky, C.~Vafa and V.~Sadov,
  ``D-Branes and Topological Field Theories,''
  Nucl.\ Phys.\ B {\bf 463}, 420 (1996)
  [arXiv:hep-th/9511222].

 \bibitem{9507090}
  M.~Cvetic and D.~Youm,
  ``Dyonic BPS saturated black holes of heterotic string on a six torus,''
  Phys.\ Rev.\ D {\bf 53}, 584 (1996)
  [arXiv:hep-th/9507090].
 
 \bibitem{9508094}
  M.~J.~Duff, J.~T.~Liu and J.~Rahmfeld,
  ``Four-Dimensional String-String-String Triality,''
  Nucl.\ Phys.\ B {\bf 459}, 125 (1996)
  [arXiv:hep-th/9508094].

\bibitem{pope}
 C.~N.~Pope,
  ``Axial Vector Anomalies And The Index Theorem In Charged 
 Schwarzschild And
  Taub - Nut Spaces,''
  Nucl.\ Phys.\ B {\bf 141}, 432 (1978).

 \bibitem{9912082}
J.~P.~Gauntlett, N.~Kim, J.~Park and P.~Yi,
 ``Monopole dynamics and BPS dyons in 
N = 2 super-Yang-Mills theories,''
  Phys.\ Rev.\ D {\bf 61}, 125012 (2000)
  [arXiv:hep-th/9912082].

\bibitem{0412287}
G.~L.~Cardoso, B.~de Wit, J.~Kappeli and T.~Mohaupt,
``Asymptotic degeneracy of dyonic N = 4 string states
and black hole
entropy,''
JHEP {\bf 0412}, 075 (2004) [arXiv:hep-th/0412287].

\end{thebibliography}
\end{document}